\documentclass[prb,reprint,superscriptaddress]{revtex4-2}

\usepackage{amsmath}
\usepackage{amsfonts}
\usepackage{graphicx}
\usepackage{bm}
\usepackage{xcolor}
\usepackage{braket}
\usepackage[normalem]{ulem}

\newcommand{\revision}{}

\def\br{{\bm x}}
\def\bR{{\bm r}}
\def\bk{{\bm k}}
\def\xR{{\bf R}}
 
\def\bA{{\bm A}} 
\def\bv{{\bm v}}
\def\bb{{\rm b}}
\def\ee{{\rm e}}
\def\fcut{{f_{\rm cut}}}
\def\fcutb{{f_{\rm cut}^\bb}}
\def\fcute{{f_{\rm cut}^\ee}}
\def\Oiii{{{\rm O}(3)}}

\def\bH{{\bm H}}
\def\bS{{\bm S}}
\def\bK{{\bm K}}
\def\xB{\Psi}

\draft % marks overfull lines with a black rule on the right

\begin{document}

%\title{An accurate and transferable data-driven environment-dependent tight binding model for aluminium} %Title of paper

\title{Equivariant analytical mapping of first principles Hamiltonians \\to accurate and transferable materials models}

\author{Liwei Zhang}
\affiliation{Department of Mathematics, University of British Columbia, 1984 Mathematics Road, Vancouver, BC, Canada, V6T 1Z2}

\author{Berk Onat}
\affiliation{Warwick Centre for Predictive Modelling, School of Engineering, University of Warwick, Coventry, CV4 7AL, United Kingdom}

\author{Genevi\`eve Dusson}
\affiliation{Laboratoire de Mathématiques, UMR CNRS 6623, Universit\'e Bourgogne Franche-Comt\'e, 16 route de Gray, 25030 Besançon, France
}

\author{\revision{Adam McSloy}}
\affiliation{Warwick Centre for Predictive Modelling, School of Engineering, University of Warwick, Coventry, CV4 7AL, United Kingdom}

\author{G. Anand}
\affiliation{Department of Metallurgy and Materials Engineering, Indian Institute of Engineering Science and Technology-Shibpur, Howrah, WB, India}

\author{Reinhard J. Maurer}
\affiliation{Department of Chemistry, University of Warwick, Coventry, CV4 7AL, United Kingdom}

\author{Christoph Ortner}
\affiliation{Department of Mathematics, University of British Columbia, 1984 Mathematics Road, Vancouver, BC, Canada, V6T 1Z2}

\author{James R. Kermode}
\email[]{J.R.Kermode@warwick.ac.uk}
\affiliation{Warwick Centre for Predictive Modelling, School of Engineering, University of Warwick, Coventry, CV4 7AL, United Kingdom}


\date{\today}

\begin{abstract}
We propose a scheme to construct predictive models for Hamiltonian matrices in atomic orbital representation from \emph{ab initio} data as a function of atomic and bond environments. The scheme goes beyond conventional tight binding descriptions as it represents the ab initio model to full order, rather than in two-centre or three-centre approximations. We achieve this by introducing an extension to the Atomic Cluster Expansion (ACE) descriptor that represents Hamiltonian matrix blocks that transform equivariantly with respect to the full rotation group. The approach produces analytical linear models for the Hamiltonian and overlap matrices. Through an application to aluminium, we demonstrate that it is possible to train models from a handful of structures computed with density functional theory, and apply them to produce accurate predictions for the electronic structure. The model generalises well and is able to predict defects accurately from only bulk training data.
\end{abstract}


\maketitle %\maketitle must follow title, authors, abstract and \pacs

%  should be done using the \cite, \ref, and \label commands
\section{Introduction} \label{sec:intro}

The availability of accurate and  highly efficient interatomic potentials is crucial for the atomistic simulation of materials phenomena with intrinsic length and time scales inaccessible to first principles electronic structure theory. Examples in materials science include failure processes such as crack propagation~\cite{Bitzek2015} and chemical dynamics at reactive surfaces~\cite{jiangDynamicsReactionsMetal2019}. The advent of machine-learning-based interatomic potentials (MLIPs) has meant that high-fidelity interatomic potentials based on Kohn-Sham Density Functional Theory (KS-DFT) and beyond have become much more widely available \cite{behlerFourGenerationsHighDimensional2021,unkeMachineLearningForce2021,deringerGaussianProcessRegression2021}. Yet, the effort to generate MLIPs that are both transferable and accurate is still significant and heavily depends on the configurational space spanned by the underlying training data set~\cite{musilPhysicsInspiredStructuralRepresentations2021}. Very few MLIPs have been reported that are able to capture different materials phases, surface terminations, and the effects of complex defects on the stability and structure of the material~\cite{deringerGaussianProcessRegression2021,Mishin2021,Behler2021}.

More importantly, MLIPs and conventional interatomic potentials fundamentally neglect explicit electronic degrees of freedom of molecules and materials thereby removing access to the simulation of observables beyond structure and stability, such as electric conductivity and  optical response, which depend on the electronic subsystem and electron-phonon coupling. While the ability to predict optical and electronic properties is desirable, the inclusion of electronic degrees of freedom will likely also benefit the transferability of MLIPs.

For decades, semi-empirical and tight-binding (TB) models of electronic structure have sought to combine the efficiency of interatomic potentials with the explicit description of electrons. A plethora of  approaches based on two-centre and three-centre integral approximations have led to established method frameworks such as the AM1 and PM3 methods \cite{Dewar1985,stewartOptimizationParametersSemiempirical1989}, the Density Functional Tight-Binding (DFTB) method~\cite{Porezag95, Elstner98}, the Sankey-Niklewski approach as implemented in the FIREBALL code \cite{Sankey1989,Lewis2001}, and the xTB approach~\cite{bannwarthGFN2xTBAccurateBroadly2019}. Unfortunately, the rigid mathematical form of the integral tabulations in most approaches means that TB parametrizations are limited in accuracy and often do not transfer beyond the materials classes for which they were originally intended.

As ML methods make inroads across a diverse range of molecular simulation workflows~\cite{westermayrPerspectiveIntegratingMachine2021}, approaches beyond MLIPs are being pursued that incorporate electronic properties. For molecules, Li et al. have proposed a neural-network-based parametrization pipeline for DFTB \cite{Li2018}, while Stoehr et al. have proposed deep tensor neural networks (DTNNs) to construct beyond-pairwise repulsion potentials~\cite{stohrAccurateManyBodyRepulsive2020}. Qiao et al. have shown that the use of symmetry-adapted atomic-orbital features can significantly improve transferability and prediction accuracy of molecular stability~\cite{qiaoOrbNetDeepLearning2020}.

%condensed phase
In the realm of condensed phase materials, the automated construction of tight-binding models from \emph{ab initio} data has been a topic of great interest as it can benefit high-throughput materials screening studies~\cite{supkaAFLOWpMinimalistApproach2017}. Most commonly, electronic structure simulations of materials are performed in non-atom-centred basis representations such as the pseudopotential plane wave framework, which is not easily amenable to the construction of TB models. 
TB Hamiltonians are typically constructed via transformation into a maximally localized Wannier function representation~\cite{garrityDatabaseWannierTightbinding2021}, which provides a compact atom-centred basis representation with local support~\cite{Marzari2012}.
It is also possible to fit Slater-Koster parameters directly to DFT calculations in a data-driven fashion~\cite{Barzdajn2021,Jenke2021}.
Materials simulations in atom-centred orbital representations as provided by, for example, the FHI-aims code~\cite{Blum2009} are becoming more common, where Wannierization is not necessary and the basis representation provided by the code is directly amenable to machine learning approaches based on local representations of atomic neighbourhoods~\cite{musilPhysicsInspiredStructuralRepresentations2021}. Examples of such representations include Behler-Parinello symmetry functions \cite{Behler2011,behlerFourGenerationsHighDimensional2021}, the SOAP descriptor~\cite{Bartok2013} or the Atomic Cluster Expansion~\cite{Drautz2019-er,Dusson2019-gn}. \revision{First efforts of direct machine learning prediction of electronic structure have been reported in literature. For example, SchNOrb~\cite{Schutt2019} is a DTNN representation of molecular mean-field electronic structure Hamiltonians, which has been used to predict Hamiltonians in local} atomic orbital and optimized effective minimal basis representations for organic molecules including up to 13 heavy atoms~\cite{Schutt2019,gasteggerDeepNeuralNetwork2020}. Hedge and Bowen~\cite{Hegde2017} employed Kernel Ridge Regression with a bispectrum representation~\cite{Bartok2010} for an analytical representation of a minimal basis DFT Hamiltonian for bulk copper and diamond. \revision{Equivariant parameterisations for molecular systems along similar lines to what we describe here have been reported, learning either from the Hamiltonian~\cite{Nigam2021-eq} or from wavefunctions and electronic densities~\cite{UnkeNeurIPS2021}. 
These works apply linear or nonlinear equivariant models, respectively, to the MD17 molecular dataset, both of which improve on the non-equivariant SchNOrb approach of Ref.~\cite{Schutt2019}. However, to our knowledge, the present work is the first to address the specific challenges of learning Hamiltonians in solid state systems.}

%\cco{Do we need to cite Ceriotti here?} \cjrk{I think we should, even though it's on arXiv and will likely change: how's this, adapted from below and compressed there.}

%SUMMARY paragraph
In this work, we present a completely data-driven approach to analytical model construction based on {\em ab initio} electronic structure theory. The model is able to faithfully represent electronic structure as a function of atomic configuration and materials composition in nonorthogonal local atomic orbital representation via the Hamiltonian and overlap matrices. This goes beyond conventional TB descriptions as it represents DFT to full order, rather than in two-centre or three-centre approximations. We achieve this by introducing an ACE descriptor to represent intraatomic onsite and interatomic offsite blocks of Hamiltonian and overlap matrices that transform equivariantly with respect to the full rotation group in three dimensions. This equivariant descriptor is integrated in an automated data-driven workflow that enables rapid parameterisation of environment dependent TB models directly from DFT data as illustrated in Fig.~1. We showcase the capabilities of this approach by predicting the band structure of bulk aluminium in different crystal systems.

\begin{figure}
    \centering
    \includegraphics[width=0.9\columnwidth]{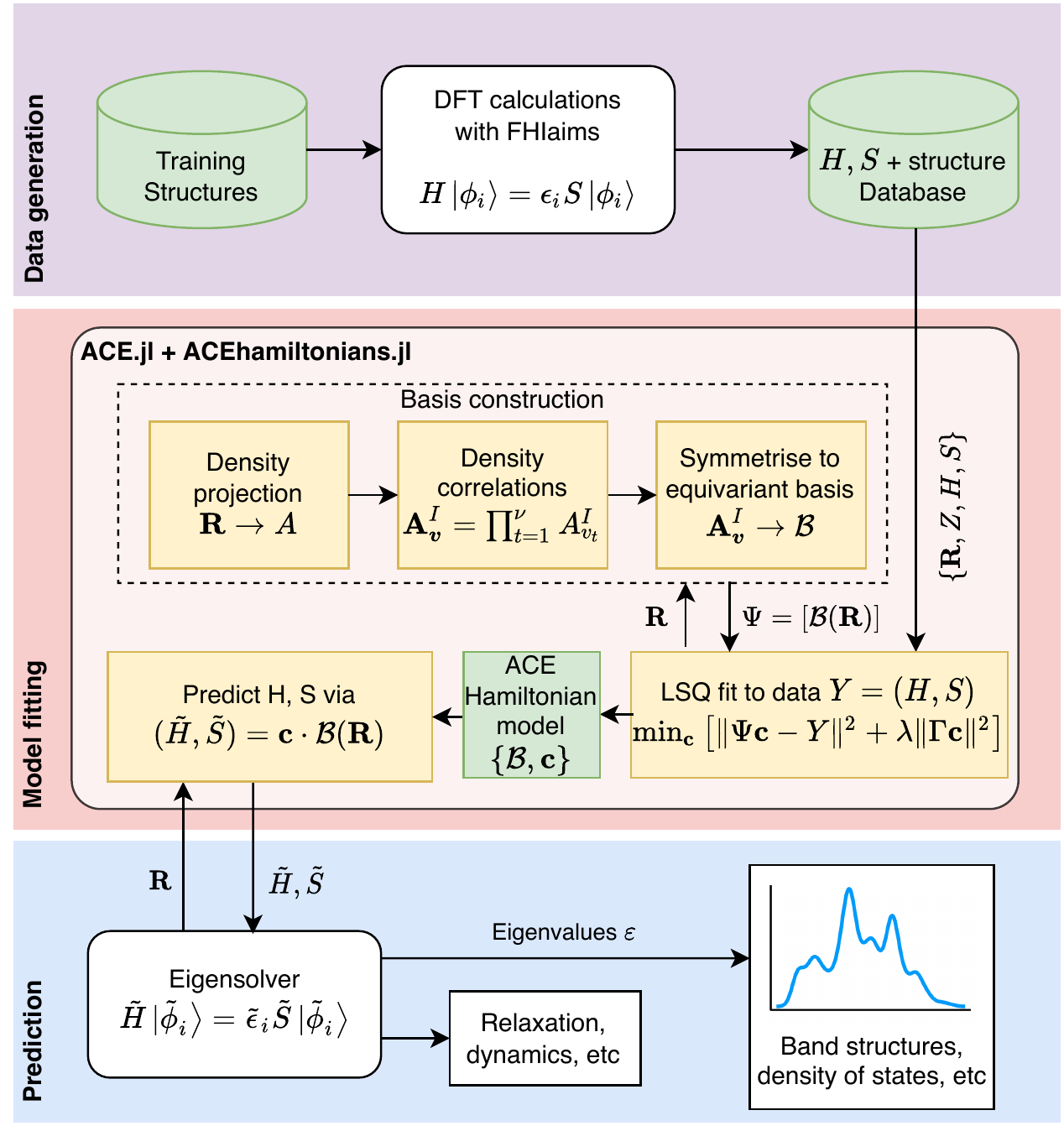}
    \caption{Schematic of the ACEhamiltonians (atomic cluster expansion for Hamiltonians) workflow, showing data generation with the FHI-aims
    electronic structure theory code, model fitting with the \texttt{ACE.jl} and \texttt{ACEhamiltonians.jl} packages, and prediction.}
    \label{fig:schematic}
\end{figure}

\section{Results} \label{sec:results}
In most electronic structure calculations the ground state of a system is obtained by solving an eigenvalue problem
\begin{equation}\label{eq:evp}
\hat{H} \psi_i = \epsilon_i\psi_i, i = 1,2,\cdots
\end{equation}
where 
\begin{equation}
    \hat{H} = - {\textstyle \frac12} \nabla^2 + V_{\rm eff}.
\end{equation}
For example, in the widely used Kohn-Sham DFT model, 
\begin{align}
    V_{\rm eff} &= V_{\rm eff}[\rho], \quad \text{where} \\
    \rho &= \sum_i f_i|\psi_i|^2,
\end{align}
and $f_i$ is the occupancy of electronic eigenstate $i$ with wave function $\psi_i$; i.e., \eqref{eq:evp} becomes a nonlinear eigenvalue problem, which is extremely computationally demanding and is usually solved by employing a Self Consistent Field (SCF) algorithm~\cite{Cances2021convergence,woodsComputingSelfconsistentField2019}.

In this paper, we are concerned with finding an analytical representation of a self-consistent Hamiltonian operator $\hat{H} = - {\textstyle \frac12} \nabla^2 + V_{\rm eff}$ in discrete basis representation. 
%In this way, we may avoid the SCF iterations and can be more flexible in dealing with many typical systems such as metals. 
% We are concerned with ``learning'' a self-consistent Hamiltonian operator of the form 
% \begin{equation}
%     \opH = - {\textstyle \frac12} \nabla^2 + V_{\rm eff},
% \end{equation}

\subsection{Hamiltonians for extended materials in atomic orbital basis representation} \label{sec:2A}
%
%\clz{Do we need a short paragraph/1-2 sentence(s) to mention the minimal basis set and how we project \eqref{eq:kse} onto it?}
%\cjrk{We're not yet actually doing that: we just use the same basis set as the DFT calculation.}
To achieve a finite basis representation, we expand the wave functions $\psi_i$ in a local nonorthogonal atom-centred basis representation
\begin{equation}\label{eq: at_basis}
    \chi_a (\br) = R_{nl}(r) Y_{lm}(\theta, \phi)
\end{equation}
%\cco{$a \equiv inlm$} 
where $a = (n, l, m; I)$ is a composite index, the spatial electron coordinate $\br$ and its components $r$, $\theta$, and $\phi$ in centrosymmetric coordinates around the atom $I$ are used. $Y_{lm}$ are spherical harmonics that define the angular dependence, and $n=0, \dots, n_{\mathrm{max}}$, $l=0, \dots, l_{\mathrm{max}}$, $m=-l_{\mathrm{max}},\dots, l_{\mathrm{max}}$ characterize the radial and angular nodal structure of the atomic orbital.
% , which we denote with a composite index $a=(n,l,m; I)$. 
The choice of $R_{nl}(r)$ varies between different types of atomic orbital basis representations and can involve linear combinations (contractions) of Gaussian functions or numerically tabulated functions. Here we choose the latter as defined in the numeric atom-centred orbital (NAO) basis employed in the FHI-aims code.~\cite{Blum2009} With this definition, we can express the overlap between basis functions and the interactions as mediated by the Hamiltonian as follows:
\begin{align}\label{eq:H}
    H_{ab} &= \braket{ \chi_{a} | \hat{H} | \chi_{b} } \qquad \text{and}  \\ 
    \label{eq:S}
    S_{ab} &=\braket{ \chi_{a} | \chi_{b} }.
\end{align}

Given a crystal-periodic structure $\xR=\{\mathbf{L}_{\kappa}, \bR_I,Z_I\}_I$ specified through a set of lattice vectors  $\mathbf{L}_{\kappa=1,2,3}$, atom positions $\bR_I$ and chemical species $Z_I$, we must consider periodic boundary conditions. As such, a Hamiltonian defined over the whole crystal volume reduces to a block diagonal Hamiltonian where each block corresponds to a vector $\bk$ in reciprocal space, which can be solved via an independent generalised eigenvalue problem:
\begin{equation}
  \mathbf{H}(\bk) \psi_{i\bk} = \epsilon_{i\bk}\mathbf{S}(\bk)\psi_{i\bk} \quad i = 1,2,\dots,
\end{equation}
where $\psi_{i\bk}$ are Bloch wave functions and $\mathbf{H}(\bk)$ and  $\mathbf{S}(\bk)$ are Hamiltonian and overlap matrices defined in terms of a discrete crystal-periodic basis. In the Methods section \ref{sec:appendix_bloch}, we show how $\mathbf{H}(\bk)$ and $\mathbf{S}(\bk)$ can be constructed at arbitrary points $\bk$ in reciprocal space from real-space representations of Hamiltonian and overlap matrices that span the full crystal volume (typically considered within a certain radius around the central unit cell). As the $\bk$-dependent matrices and the solution of the set of generalised eigenvalues completely follow from the real-space $\bH$ and $\bS$ in \eqref{eq:H} and \eqref{eq:S}, we will go on to develop a representation for those two matrix quantities as a function of the structure $\xR$.

Recall that $\hat{H} = - \frac12 \nabla^2 + V_{\rm eff}$. The effective potential $V_{\rm eff}$ is not only a function of the spatial electron coordinate $\br$ but also of the entire atomic structure, i.e., one should think of 
\begin{equation}
    V_{\rm eff} = V_{\rm eff}(\br; \xR ).
\end{equation}
For example, in KS-DFT, this dependence arises due to the dependence of $V_{\rm eff}$ on the self-consistent electron density.
Our aim will be to construct a general regression scheme for the discretised Hamiltonian exploiting three fundamental, general properties of $\hat{H}$ and in particular $V_{\rm eff}$: 
(i) near-sightedness of electronic structure; 
(ii) smoothness under changes in the atomic structure; and 
(iii) equivariance of the Hamiltonian. We will discuss in the next section how these properties are to be exploited in the parameterisation. 

In preparation, we first make (iii) more precise: 
let $Q \in \Oiii$ denote an isometry (rotation and reflection) and $Q\xR = \{\mathbf{L}_{\kappa}, Q \bR_I,Z_I\}_I$ (where we also rotate the cell). Further, let $\bH_{IJ} = \bH_{IJ}(\xR)$ denote the Hamiltonian block corresponding to interactions between orbitals centered at sites $I$ and $J$. It is then straightforward to deduce that 
\begin{equation} \label{eq:equiv_H}
    \bH_{IJ}( Q \xR ) = D(Q)^* \bH_{IJ}(\xR) D(Q),
\end{equation}
where $D(Q)$ is a block-Wigner-D matrix,
\begin{equation}
    D(Q) = \textup{Diag}(D^{l_1}(Q), D^{l_2}(Q), \cdots),
\end{equation}
and $(l_1, l_2, \dots)$ specify the types of orbitals at each site. More details can be found in Methods section \ref{sec:appendix_equivariant}. Since the focus of the present work is on elemental metallic systems we ignore chemical species information entirely in the present work; this will be addressed in the future either directly as is done for ACE interatomic potentials~\cite{Lysogorskiy2021} or using compressed species information~\cite{Willatt2018-ao,Nigam2020-re}.

Crucially, there are only two distinct functional relationships that must be ``learned'' in order to represent the entire Hamiltonian: one for off-site blocks that represent interactions between orbitals centered at two different atoms and one for on-site blocks representing interactions of orbitals at the same atom. More precisely, the translation invariance and permutation equivariance of the Hamiltonian imply that 
\begin{equation} \label{eq:Honoff}
    \begin{split} 
        \bH_{II} &= \bH_{\text{on}}(\xR_I), \qquad \text{and} \\ 
        \bH_{IJ} &= \bH_{\rm off}{\revision{(\xR_{IJ})}}, 
    \end{split}
\end{equation}
%\clz{should we use double-subscript for $\bR_{IJ}$ (i.e., $\bR_{_{IJ}}$) and also for all other small r with subscript?}
where $\xR_I$ denotes the {\em atomic environment} of atom $I$ and \revision{ $\xR_{IJ}$ the {\em bond environment} of (multiple) bonds between the two atoms $i, j$ which also contains the position of bonds.} These environments are defined as follows: 
\begin{align}
    \xR_I &:= \big\{ \bR_{IK} \,|\, K \neq I \big\}, \qquad \text{and} \nonumber \\ 
    \xR_{IJ} &:= \revision{\big\{\bR_{IJ}; \{\bR_K - {\textstyle \frac12 (\bR_I + \bR_J)} \,|\, K \neq I,J\} \big\},}
\end{align}
{\revision{ where $\bR_{IJ} = \bR_I - \bR_J$.}}
%\clz{Here (from $\xR$ to $\xR_I, \xR_{IJ}$), we seemed to default that all $Z_I$ are the same but didn't mention that.} 
In the above definitions, the index $K$ runs over all unit cells $N$ within the crystal volume. According to \eqref{eq:equiv_H} the functions $\bH_{\text{on}}$ and $\bH_{\rm off}$ are equivariant in the sense that 
\begin{equation}\label{eq:cov}
   \bH_{\text{on/off}}(Q\xR) = D(Q)^* \bH_{\text{on/off}}(\xR) D(Q). 
\end{equation}
Translation invariance is now built into the dependence of $\bH_{\rm on/off}$ on relative positions only, while permutation equivariance of $\bH$ is built into \eqref{eq:Honoff}.

Several simplifications apply for the treatment of the overlap matrix. For each atom we choose a set of basis functions $\chi$ that are orthogonal, which means that the on-site blocks $\bS_{II}$ are identity matrices. 
%\clz{should that be further specified, e.g., are identities/multiple of identities? Or else it still needs to be fitted.}.
The off-site blocks follow the same symmetry as the Hamiltonian off-site blocks.

\subsection{Parameterisation}

\begin{figure*}
    \centering
    \includegraphics[width=0.8\textwidth]{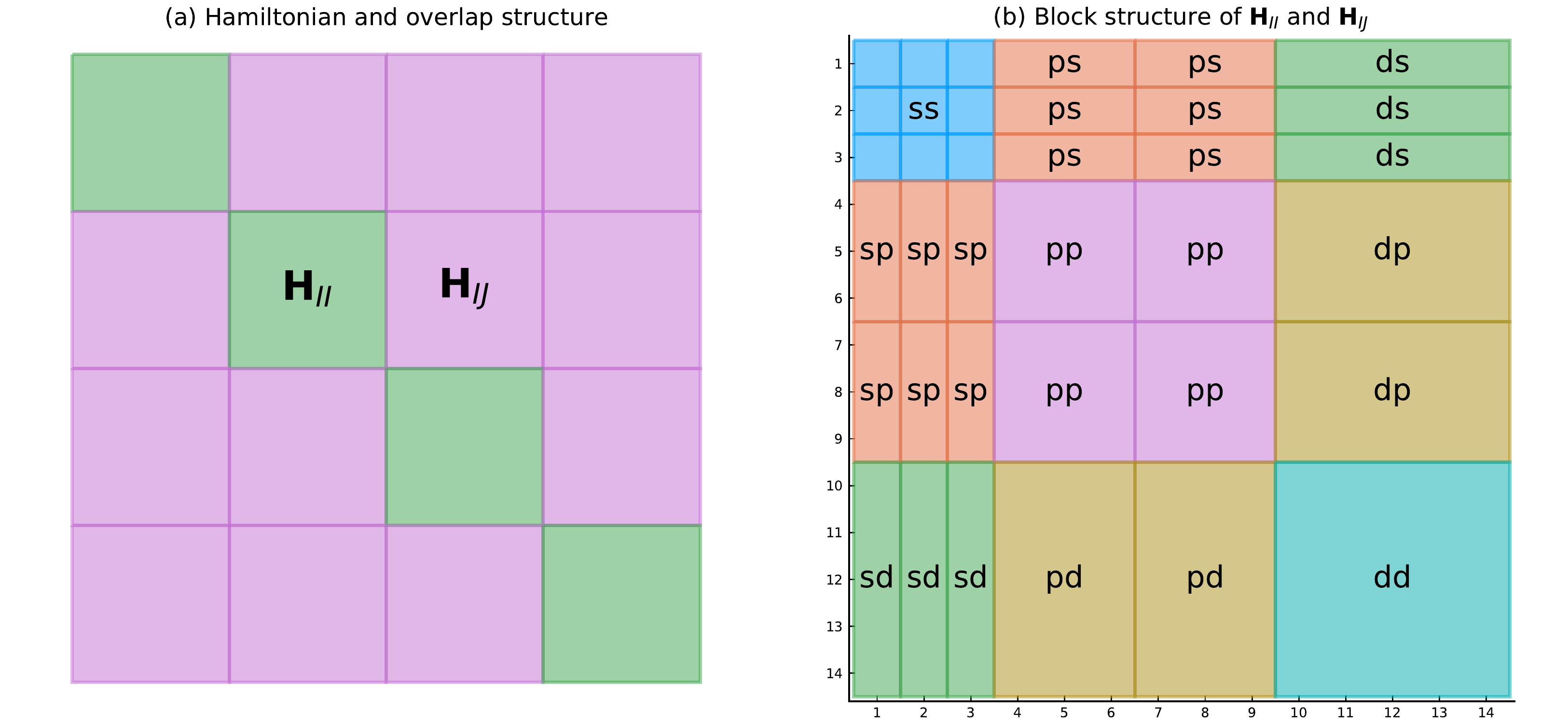}
    \caption{Block structure and atomic orbital subblocks in the Hamiltonian and overlap matrices used in our models. Each block within panel (a) is a $14\times14$ matrix with the atomic orbital structure $\bH_{IJ}$ shown in panel (b). Blocks coloured green in (a) are onsite blocks, while those shown in purple are offsite blocks. Note that the onsite $\bH_{II}$ are self-adjoint and hence, e.g., only one of the $ps$ and $sp$ blocks needs to be fitted.}
    \label{fig:HS_schematic}
\end{figure*}

We parameterise the real-space Hamiltonian and overlap matrix blocks $\bH_{\rm on}, \bH_{\rm off}$ and $\bS_{\rm off}$ using an equivariant ACE basis~\cite{Drautz2019-er,Dusson2019-gn,Drautz2020-mg}. Similar techniques have previously been proposed in other contexts~\cite{Grisafi2019-in,Nigam2020-re,Nigam2021-eq}. In this section, we present a general outline of the ideas, making certain choices of approximation parameters concrete in the Methods \S \ref{sec:parameter_choices}.

We denote the parameterised Hamiltonian and overlap by $\tilde{\bH}, \tilde{\bS}$. For the sake of simplicity we focus the presentation on $\tilde{\bH}$ and remark on the relevant modification for $\tilde{\bS}$ at the end. All  procedures are straightforward to generalise for multiple species with the only effect being an increased number of $\tilde{\bH}$ and $\tilde{\bS}$ blocks that have to be considered as element combinations increase. In the present case, $\tilde{\bH}_{\rm on}$ is invariant under permutations of $\xR_I$ and $\tilde{\bH}_{\rm off}$ is invariant under permutations of $\xR_{IJ}$. 
Both can therefore be parameterised by the ACE model. Here, we closely follow the procedures introduced in Refs.~\cite{Drautz2020-mg,Dusson2019-gn,Lysogorskiy2021}.

{\it 1. Parameterisation of $\bH_{\rm on}$: } We start by choosing a {\it one-particle} basis, 
\begin{equation} \label{eq:one-particle-on}
    \phi_v(\br) := \phi^{\rm on}_{nlm}(\br) := P_{nl}(r) Y_{lm}(\hat{\br}) \fcut(r)
\end{equation}
where  $\br = r\hat{\br}$ and we have identified the composite index $v \equiv (nlm)$. The radial cutoff or envelope function $\fcut(r)$ ensures that only interactions of nearby atoms are taken into account, exploiting the near-sightedness of electronic structure. 

Given the one-particle basis we can form the density projection and projected $\nu$-correlations (product basis), 
%\cgd{should it be $ A_v^I := \sum_{j \neq i} \phi_v(\bR_{ij})$ just below ? Or $A_v^I &:= \sum_{J \neq I} \phi_v(\bR_{IJ})$?}
\begin{align}
    A_v^I &:= \sum_{J \neq I} \phi_v(\bR_{IJ}), \\ 
    \bA_{\bv}^I &:= \prod_{t = 1}^\nu A_{v_t}^I 
        \qquad \text{for } \bv = (v^1, \dots, v^\nu), \nu = 1, 2, \dots.
\end{align}
The $\bA_{\bv}^I$ form a complete basis of permutation-invariant (PI) polynomials, hence we can approximate 
\begin{equation} \label{eq:Hon_pi}
    \bH_{II} = \bH_{\rm on}(\xR_{I}) \approx \tilde{\bH}_{\rm on}^{\rm PI}(\xR_I) = \sum_{\bv} C_{\bv} \bA_{\bv}^I,
\end{equation}
where $\bA_{\bv}^I$ are scalar and the parameters $C_\bv = (C_\bv^{\alpha_1 \alpha_2})_{\alpha_1, \alpha_2 = 1}^{N_{\rm orb}}$ have the same dimensionality as $\bH_{II}$ i.e., $N_{\rm orb} \times N_{\rm orb}$ (recall that $\bH_{II}$ denotes the onsite Hamiltonian block corresponding to orbitals centered at atom $I$). The summation over $\bv$ will be restricted to a finite set, the choice of which is a crucial aspect of the model accuracy; cf. \S~\ref{sec:parameter_choices}.

The expansion \eqref{eq:Hon_pi} incorporates translation and permutation invariance but not yet the $\Oiii$-equivariance~\eqref{eq:equiv_H}. Following the general ACE construction~\cite{Dusson2019-gn} we can achieve this by simply averaging the representation over the group $\Oiii$, i.e., 
\begin{equation}
    \tilde{\bH}_{\rm on}(\xR_I) = -\!\!\!\!-\hspace{-4.5mm}\int_{\Oiii} D(Q) \tilde{\bH}_{\rm on}^{\rm PI}(Q\xR_I) D(Q)^* dQ,
\end{equation}
In step 4. we will review how this integration is explicitly resolved. 

{\it 2. Parameterisation of $\bH_{\rm off}$: } The procedure for parameterising $\bH_{\rm off}$ is similar to that of $\bH_{\rm on}$, the main difference being that the presence of a bond rather than a site changes the permutation-invariance. Specifically, we now need to define one-particle basis functions for the bond variable and for the environment variables
\begin{equation}  \label{eq:one-particle-off}
   \begin{split}
   \phi_{nlm}^\bb(\bR_{IJ}) &= P_{nl}^\bb(r_{IJ}) Y_{lm}(\hat{\bR}_{IJ}) \fcutb(r_{IJ}), \\ 
   \phi_{nlm}^\ee(\bR_{IJ,K}) &= P_{nl}^\ee(r_{IJ,K}) Y_{lm}(\hat{\bR}_{IJ,K}) \fcute(\bR_{IJ,K}, \bR_{IJ}).
%   \\ 
%   \phi^{\text{off}}_{(n,l,m,b)}(\sigma) &= \phi_{nlm}(\br)f_{r_{\text{cut}}}(|\br|)\delta_{b,symbol}, \\
%   \phi^{\text{off}}_{(n,l,m,e)}(\sigma) &= \phi_{nlm}(\br)f_{\mathcal{E}}(\br)\delta_{e,symbol}
   \end{split}
\end{equation}
where $\bR_{IJ} = r_{IJ}\hat{\bR}_{IJ}$ and $\bR_{IJ,K} := \bR_K - \frac12(\bR_I + \bR_J)$.
Note in particular that the cutoff function for the environment, $\fcute$, no longer depends only on the radius but may be more general: we require only that $\fcute(\bR_{IJ,K}, \bR_{IJ})$ is invariant under joint rotation of both arguments which allows, e.g., ellipsoidal or cylindrical cutoff geometries.

The density projection for the bond environment $\xR_{IJ}$ is now given by 
%\clz{ And similarly, do we need double superscript for $A^{IJ}$ (i.e., $A^{^{IJ}}$) }
\begin{equation}
    A^{IJ}_v := \sum_{K \neq I, J} \phi_v^\ee(\bR_{IJ,K}), 
\end{equation}
and the product basis becomes 
\begin{equation}
    \bA_{\bv}^{IJ} := \phi^\bb_{v^0}(\bR_{IJ}) \cdot \prod_{t = 1}^\nu A^{IJ}_{v^t}, 
\end{equation}
for $\bv = (v^0, v^1, \dots, v^\nu)$, with $\nu = 0, 1, 2, \dots$ the correlation order of the bond environment. As in the on-site case, the $A^{IJ}_{\bv}$ form a complete basis of polynomials that are invariant under permutations of $\xR_{IJ}$ and we may therefore approximate
\begin{equation}
    \bH_{IJ} = \bH_{\rm off} \approx \tilde{\bH}_{\rm off}^{\rm PI}\revision{(\xR_{IJ})}
    := \sum_{\bv} C_\bv \bA_{\bv}^{IJ}.
\end{equation}
which we finally symmetrize to obtain also the $O(3)$-equivariance, 
\begin{equation}
    \tilde{\bH}_{\rm off}\revision{(\xR_{IJ})} 
    := -\!\!\!\!-\hspace{-4.5mm}\int_{\Oiii} \!\!\!\!\! D(Q) \tilde{\bH}_{\rm off}^{\rm PI}\revision{(Q\xR_{IJ})} D(Q)^* dQ.
\end{equation}

{\it 3. Parameterisation of $\bS_{\rm off}$: } The environment-dependence of $\bH_{\rm off}$ enters only through the effective potential $V_{\rm eff}$ which is not present in the overlap matrix definition. Therefore, we simply parameterise $\bS_{\rm off}$ by 
\begin{equation}
    \tilde{\bS}_{\rm off}(\bR_{IJ}) := -\!\!\!\!-\hspace{-4.5mm}\int_{\Oiii} \!\!\!\!\! D(Q) \bigg[ 
            \sum_v C_v \phi^\bb_{v}(Q \bR_{IJ}) \bigg] D(Q)^* \, dQ.
\end{equation}
This is formally equivalent to a Slater Koster representation of two-centre integrals~\cite{Slater54}, which is exact in the case of the overlap. For our ACE parameterisation, this means that we only need to use correlation order $\nu = 0$, i.e. no environment-dependence of the bond integral needs to be considered. 
%\cco{comment on the fact that we could just do this explicitly if we wanted? So why don't we? Too much work? Feels a bit weird. Maybe we need to say that this is just Slater-Koster but written in a weird and roundabout way? }

{\it 4. Recursive symmetrisation: } In all three cases $\tilde{\bH}_{\rm on}, \tilde{\bH}_{\rm off}, \tilde{\bS}_{\rm off}$ we have reduced the parameterisation to an integral over the symmetry group $O(3)$, i.e.,
\begin{equation} \label{eq:Ktilde}
    \tilde{\bK}(\xR_\bullet) = 
    -\!\!\!\!-\hspace{-4.5mm}\int_{\Oiii} \!\!\!\!\! D(Q)
    \bigg[ \sum_{\bv} C_{\bv} \bA^\bullet_{\bv}(Q \xR_\bullet) \bigg] D(Q)^*,
\end{equation}
where $\tilde{\bK}$ denotes one of the three model components $\tilde{\bH}_{\rm on}, \tilde{\bH}_{\rm off}, \tilde{\bS}_{\rm off}$ and $\xR_\bullet$ denotes an atom environment $\xR_I$ or bond environment $\xR_{IJ}$. In particular, for off-site overlap $\bS_{\rm off}$, 
\begin{equation}
    \bA^{IJ}_{\bv}(\xR_{IJ}) = \phi^\bb_{v}(\bR_{IJ}).
\end{equation}
In order to make our description clearer, we denote $\bv \equiv {\boldsymbol{nlm}}$ with $\boldsymbol{n,l,m}$ being the lists of corresponding indices in $\bA^{IJ}_{\bv}$. Thus, we can deduce that

\begin{equation}
   \bA^\bullet_{\boldsymbol{nlm}}(Q\xR_\bullet)= \sum_{\boldsymbol{\mu}}{\bm D}^{{\bm l}}_{\boldsymbol{\mu m}}(Q) \bA^\bullet_{\boldsymbol{nl\mu}}(\xR_\bullet),
\end{equation}
where ${\bm D}^{{\bm l}}_{\boldsymbol{\mu m}}(Q) = \prod_{t} D^{l_t}_{\mu_t m_t}(Q)$ since the angular dependence of the one-particle basis functions in all cases is in terms of spherical harmonics $Y_{lm}$.
Furthermore, we write 
\begin{equation}
    C_\bv = \sum_{\alpha, \beta = 1}^{N_{\rm orb}} c_{\bv}^{\alpha\beta}  E^{\alpha\beta},
\end{equation}
where $E^{\alpha\beta} \in \mathbb{R}^{N_{\rm orb} \times N_{\rm orb}}$ with $E^{\alpha\beta}_{\alpha'\beta'} = \delta_{\alpha\alpha'}\delta_{\beta\beta'}$. Inserting these two identities into \eqref{eq:Ktilde} yields
\begin{equation}  \label{eq:Ktilde_sym_overcomplete}
   \begin{split}
   \tilde{\bK}(\xR_\bullet) 
   &= \sum_{\bm n, \bm l, \bm m,\alpha,\beta} c^{\alpha\beta}_{\bv}
    \sum_{\boldsymbol{\mu}}{\mathcal{U}}^{\alpha\beta}_{{\bm l \boldsymbol{\mu} \bm m}}\bA^\bullet_{\bm n \bm l \boldsymbol{\mu}}(\xR_\bullet) \\
   &=:\sum_{\boldsymbol{n,l,m},\alpha,\beta}c^{\alpha\beta}_{\boldsymbol{nlm}}\mathcal{B}^{\alpha\beta}_{\boldsymbol{nlm}}(\xR_\bullet),
   \end{split}
\end{equation}
   where the ``generalized coupling coefficients'' are given by 
\begin{equation}
    {\mathcal{U}}^{\alpha\beta}_{{\bm l\boldsymbol{\mu}m}} = -\!\!\!\!-\hspace{-4.5mm}\int_{\Oiii} \!\!\!\!\! {\bm D}^{\bf l}_{\boldsymbol{\mu}\bf m}(Q)D(Q) E^{\alpha\beta}D(Q)^* dQ.
\end{equation}
Their definition involves an integral over products of Wigner-D matrices which can be precomputed explicitly (i.e., without need for quadrature which would incur a discretisation error) using the recursion proposed by Dusson \emph{et al.}~\cite{Dusson2019-gn} and independently by Nigam \emph{et al.}~\cite{Nigam2021-eq}. 
% This is a key step where our approach differs from the NICE framework\cco{cite} used by \cite{Nigam2021-eq}: it is formally equivalent to the ACE framework, but takes a different algorithmic approach resolving the recursion during evaluation rather than via precomputation.\cbo{The difference of our work from theirs is not clear to me here. Shouldn't we give more details on the differences and make them clear and bold?}

Note that \eqref{eq:Ktilde_sym_overcomplete} parameterises $\tilde{K}$ in terms of the scalar parameters $c_{\bv}^{\alpha\beta}$, while the basis functions are now matrix-valued, 
\begin{equation}
    \mathcal{B}^{\alpha\beta}_{\boldsymbol{nlm}}(\xR_\bullet) = \sum_{\boldsymbol{\mu}}{\mathcal{U}}^{\alpha\beta}_{{\bf l\boldsymbol{\mu}m}}\bA^\bullet_{\bm n \bm l\boldsymbol{\mu}}(\xR_\bullet).
\end{equation}
Since the coupling coefficients $\mathcal{U}$ are extremely sparse, the operation to obtain $\mathcal{B}$ from $\bA^\bullet$ is relatively cheap.
   
Due to the coupling, the basis $\mathcal{B}_{\boldsymbol{nlm}}^{\alpha\beta}$ is normally overcomplete. This linear dependence arises exactly within fixed $\boldsymbol{nl}$ blocks. In a straightforward adaption of the general procedures outlined by Dusson \emph{et al.} \cite{Dusson2019-gn} we use elementary linear algebra techniques to reduce the basis in a block-by-block fashion by constructing reduced coupling coefficients $\mathcal{U}^{\boldsymbol{nl}}_{k\bm \mu}$ and defining 
\begin{equation}
   \mathcal{B}_{\boldsymbol{nl}k}(\xR_\bullet) := 
    \sum_{\boldsymbol{\mu}}{\mathcal{U}^{\boldsymbol{nl}}_{k\bm \mu}} \bA^\bullet_{\bm n \bm l \boldsymbol{\mu}}(\xR_\bullet).
\end{equation}

In summary, after dropping the detailed multi-index notation and replacing it with a simple enumeration of the basis, we obtain {\em linear models} for 
\begin{align}
    \tilde{\bH}_{\rm on} &:= {\bf c}^{\rm on} \cdot \mathcal{B}^{\rm on}, \\ 
    \tilde{\bH}_{\rm off} &:= {\bf c}^{\rm off} \cdot \tilde{\mathcal{B}^{\rm off}}, \\ 
    \tilde{\bS}_{\rm off} &:= {\bf c}^{\rm S} \cdot \tilde{\mathcal{B}^{\rm S}}, \\ 
\end{align}
all of which inherit exactly the translation and permutation invariance as well as $\Oiii$-equivariance of $\bH_{\rm on}, \bH_{\rm off}, \bS_{\rm off}$. In the limit of infinite basis size and infinite cutoff radius these models can (in principle) be converged to within arbitrary accuracy. In this sense, they are {\em universal}. After imposing the symmetries outlined above we still need to ensure self-adjointness of the assembled Hamiltonian and overlap operators which we achieve by simply substituting $\tilde{\bH} \leftarrow \frac12(\tilde{\bH}+\tilde{\bH}^\ast)$, and analogously for the overlap.

% {\revision The imposed symmetries outlined above are all in-block symmetries. As such, the above parameterisations fail to ensure self-adjointness of both predicted $\tilde{\bH}$ and $\tilde{\bS}$ though their self-adjointness will eventually be caught with the increase of both training set and number of basis. Hence, we explicitly let
% \begin{equation}
% \tilde{\bH} \leftarrow \frac{\tilde{\bH}+\tilde{\bH'}}{2} \text{ and } \tilde{\bS} \leftarrow \frac{\tilde{\bS}+\tilde{\bS'}}{2},
% \end{equation}
% as a post-processing.
% }

% After determining the model parameters in the equivariant basis we can contract them with the coupling coefficients to obtain a more compact representation that we use for evaluation: 
%   \begin{align}
%         \tilde{\bK}
%         &= \sum_\tau c_\tau \mathcal{B}_\tau  
%         = \sum_\tau c_\tau \sum_{\bm v} \mathcal{U}^\tau_{\bm v} \bA_{\bm v} \\ 
%         &= \sum_{\bm v}  \bigg[ \sum_\tau \mathcal{U}^\tau_{\bm v} \bigg] \bA_{\bm v} 
%         =: \sum_{\bm v} C_\bv \bA_\bv.
%   \end{align}
% where we have again reverted to a scalar-valued basis $\bA_\bv$ and matrix-valued coefficients $C_\bv$. In this representation, the equivariance is encoded in the coefficients rather than in the basis. 

\subsection{Validation}

\begin{figure*}
    \centering
    \includegraphics[width=\textwidth]{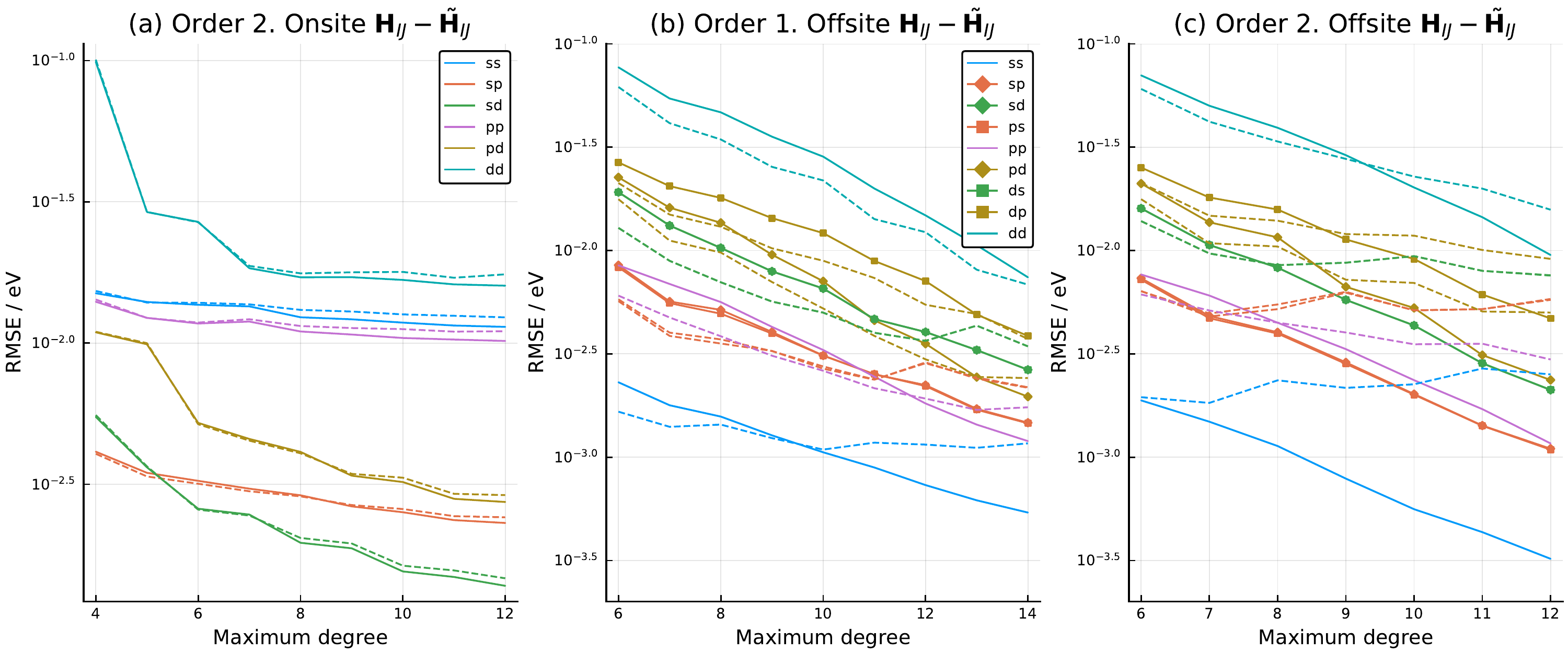}
\caption{Convergence of Hamiltonian and overlap blocks with respect to the order and maximum degree of the ACE basis set.  (a) Onsite Hamiltonian blocks $\bH_{II}$ fitted with order 2 models of varying maximum degree. (b) Offsite Hamiltonian blocks with order 1 ACE models.(c) Offsite Hamiltonian blocks with order 2 ACE models. In all plots solid lines show errors on training data and dashed lines errors on test data. Colours match the block structure of Fig.2. Note the distinct markers that distinguish the non-adjoint entries in the offsite Hamiltonian and overlap blocks.}
    \label{fig:HS_convergence}
\end{figure*}

We generated DFT data for FCC and BCC aluminium, and followed the procedure outlined above to construct ACE models for the Hamiltonian and overlap using several choices of basis sets. Full details of data generation, parameter estimation and prediction procedures are given in Methods (section  \ref{sec:methods}).

The ACE basis sets need to be carefully chosen for a particular application. The larger the basis, the higher the achievable accuracy, but larger basis sets also carry a risk of loss of transferability through overfitting.
%
%For a given orbital angular momentum $l$, each basis set is defined by two parameters: the correlation order $\nu$ and the maximum polynomial degree $n_\mathrm{max}$ used in the radial basis functions $P_{nl}(r)$ of \eqref{eq:one-particle-on} and \eqref{eq:one-particle-off}.
Each basis set is defined by three parameters: the correlation order $\nu$ and the maximum polynomial degrees $n_\mathrm{max},  l_\mathrm{max}$ used in both the radial basis functions $P_{nl}(r)$ and the angular basis function $Y_{lm}(\hat{\br})$ of \eqref{eq:one-particle-on} and \eqref{eq:one-particle-off}. In all our tests, the polynomial degrees are truncated in the manner of total degree, i.e., we let $n+l\le d_\mathrm{max}$ for a given $d_\mathrm{max}$.
For the onsite models, the body order is one more than the correlation order, i.e. $\nu=1$ corresponds to two body and $\nu=2$ to three body, while for the offsite models the body order is two more than the correlation order (since each term in the body order expansion depends on the bond in addition to $\nu$ particles from the environment). The offsite model has further flexibility in that one can choose different $d_\mathrm{max}$ for bond and environment, say, $d^{\rm b}_\mathrm{max}$ and $d^{\rm e}_\mathrm{max}$. To avoid overemphasising the impact of environment, we set $d^{\rm e}_\mathrm{max} = \lceil d^{\rm b}_\mathrm{max}/2 \rceil$ in our implementation.

We tested the accuracy of the fitted Hamiltonian and overlap matrices using different choices of these basis set parameters. The results are illustrated in Fig.~3. For the onsite blocks $\bH_{II}$, we can obtain accurate and transferable results for all sub-blocks with correlation order $\nu=2$ (body order 3), with no significant overfitting as can be seen from the close agreement of prediction accuracies on the training and test datasets in Fig.~3(a). The largest errors are on the $dd$ subblock, which also has the largest matrix entries; the RMSE of $\sim10$~meV on this sub-block corresponds to a $\sim$2\% relative error. 

For offsite blocks $\bH_{IJ}$ we considered models with correlation orders of both $\nu=1$ (body order 3), Fig.~3(b), and $\nu=2$ (body order 4), Fig.~3(c).
Both approaches show good convergence in the accuracy on the training set as the maximum degree is increased. However, for sub-blocks that include interaction with $s$ orbitals, we observe that overfitting occurs at lower degrees for the order 2 models than for the order 1 case. We speculate that this might result from the higher order basis sets providing too much flexibility for functions that have relatively simple functional behaviour. Since $s$ orbitals have no intrinsic rotational dependence, all rotational equivariance behaviour in $sp$ and $sd$ sub-blocks comes from how the $p$ or $d$ orbitals are positioned with respect to the environment.

We find the correlation order 1 models provide sufficient accuracy, in fact closely comparable to that of the order 2 models on the training set, so to avoid issues of overfitting we use order 1 only for $\bH_{IJ}$, and also limit the maximum polynomial degree for individual sub-blocks as discussed in more detail in \S~\ref{sec:optimal-model} below.

As expected from the lack of environment dependence, the offsite overlap $\bS_{IJ}$ is very well reproduced at correlation order 0 (body order 2), with a RMSE of $10^{-4}$. We do not observe any over-fitting for the offsite overlap so we fixed the maximum polynomial degree for $\bS_{IJ}$ at 16, the highest value we tried.

\subsection{Cross-validation and Model Selection} \label{sec:optimal-model}

\begin{table}[]
    \centering
    \begin{tabular}{lcccc}
        \multicolumn{5}{l}{Onsite Hamiltonian $\bH_{II}$} \\
        \hline
        \hline
        Correlation order $\nu$ & \multicolumn{4}{c}{2} \\
        Cutoff radius $r_\mathrm{cut}$ & \multicolumn{4}{c}{10~\AA} \\
        Maximum polynomial degree $d_\mathrm{max}$ &  \multicolumn{4}{c}{9}  \\
        Regularisation $\lambda$ &  \multicolumn{4}{c}{$10^{-7}$} \\
        & & & & \\
        \multicolumn{5}{l}{Offsite Hamiltonian $\bH_{IJ}$} \\
        \hline
        \hline
        Correlation order $\nu$ & \multicolumn{4}{c}{1} \\
        Bond cutoff radius $r^{\rm b}_\mathrm{cut}$ & \multicolumn{4}{c}{10~\AA} \\
        Env. cutoff radius $r^{\rm e}_\mathrm{cut}$ & \multicolumn{4}{c}{5~\AA} \\
        Env. cutoff radius $z^{\rm e}_\mathrm{cut}$ & \multicolumn{4}{c}{5~\AA} \\
        Maximum polynomial degree $d^{\rm b}_\mathrm{max}$ &       & 14  & 14  & 14 \\
                                      & $ss$  & 14  & 14  & 14 \\
                                      &       & 14  & 14  & 9 \\
                                      \cline{2-5}
                                      & $sp$  & 14  & 14 & 12 \\
                                      &       & 14 & 14 & 10 \\
                                      \cline{2-5}
                                      & $sd$  & 14 & 14 & 11 \\
                                      \cline{2-5}
                                      & $pp$  & 13 & 13 \\
                                      &       & 13 & 13 \\
                                      \cline{2-4}
                                      & $pd$  & 14 & 14 \\
                                      \cline{2-4}
                                      & $dd$  & 14 \\
        Regularisation $\lambda$ &  \multicolumn{4}{c}{$10^{-7}$} \\
        & & & & \\
        \multicolumn{5}{l}{Offsite overlap $\bS_{IJ}$} \\
        \hline
        \hline
        Correlation order $\nu$ & \multicolumn{4}{c}{0} \\
        Cutoff radius $r_\mathrm{cut}$ & \multicolumn{4}{c}{10~\AA} \\
        Maximum polynomial degree $d_\mathrm{max}$ &  \multicolumn{4}{c}{16}  \\
        Regularisation $\lambda$ &  \multicolumn{4}{c}{$10^{-7}$}
    \end{tabular}
    \caption{ACE basis set parameters for our optimised models for $\bH_{II}$, $\bH_{IJ}$ and  $\bS_{IJ}$. Maximum polynomial degree can be specified independently for each component model shown in Fig.2. The maximum polynomial degrees for the adjoint blocks $ps$, $ds$ and $dp$ of $\bH_{IJ}$ are the transposes of those shown for $sp$, $sd$ and $pd$, respectively.}
    \label{tab:optimal-model}
\end{table}

To eliminate overfitting we used the cross-validation results illustrated in Fig.~3 to select a customised basis set for each sub-block, as set out in Table \ref{tab:optimal-model}. Note that the maximum polynomial degree can be chosen for each individual sub-block model shown in the schematic in Fig.~1, i.e. there are 9 $ss$ models, $3\times2 =6$ $sp$ models, and $2 \times 2 =4$ $pp$ models.
For the $3\times3=9$ $ss$ sub-blocks of the offsite Hamiltonian we found it necessary to reduce the degree only for the $3s-3s$ entry, which arises from the fact that the FHI-aims basis set features two $s$ orbitals in the valence shell of Al.

We used our optimised model to predict the Hamiltonian and overlap for the FCC and BCC equilibrium crystal geometries. These were not included in the training set, which comprises only perturbed structures from molecular dynamics, so can be viewed as a test of its transferability. The magnitudes and associated errors in the onsite and one of the nearest-neighbour offsite blocks of the Hamiltonian matrix are illustrated in Fig.~4 for the FCC case; BCC results are of comparable accuracy. These results demonstrate the correct equivariance of the predictions with matrix entries, i.e. entries which should be zero by symmetry being correctly captured. Comparing the upper and lower panels also illustrates that the errors are always orders of magnitude smaller than the corresponding magnitudes, ensuring that the relative error is well controlled (typically $\sim$1\% or less).

\begin{figure}
    \centering
    \includegraphics[width=\columnwidth]{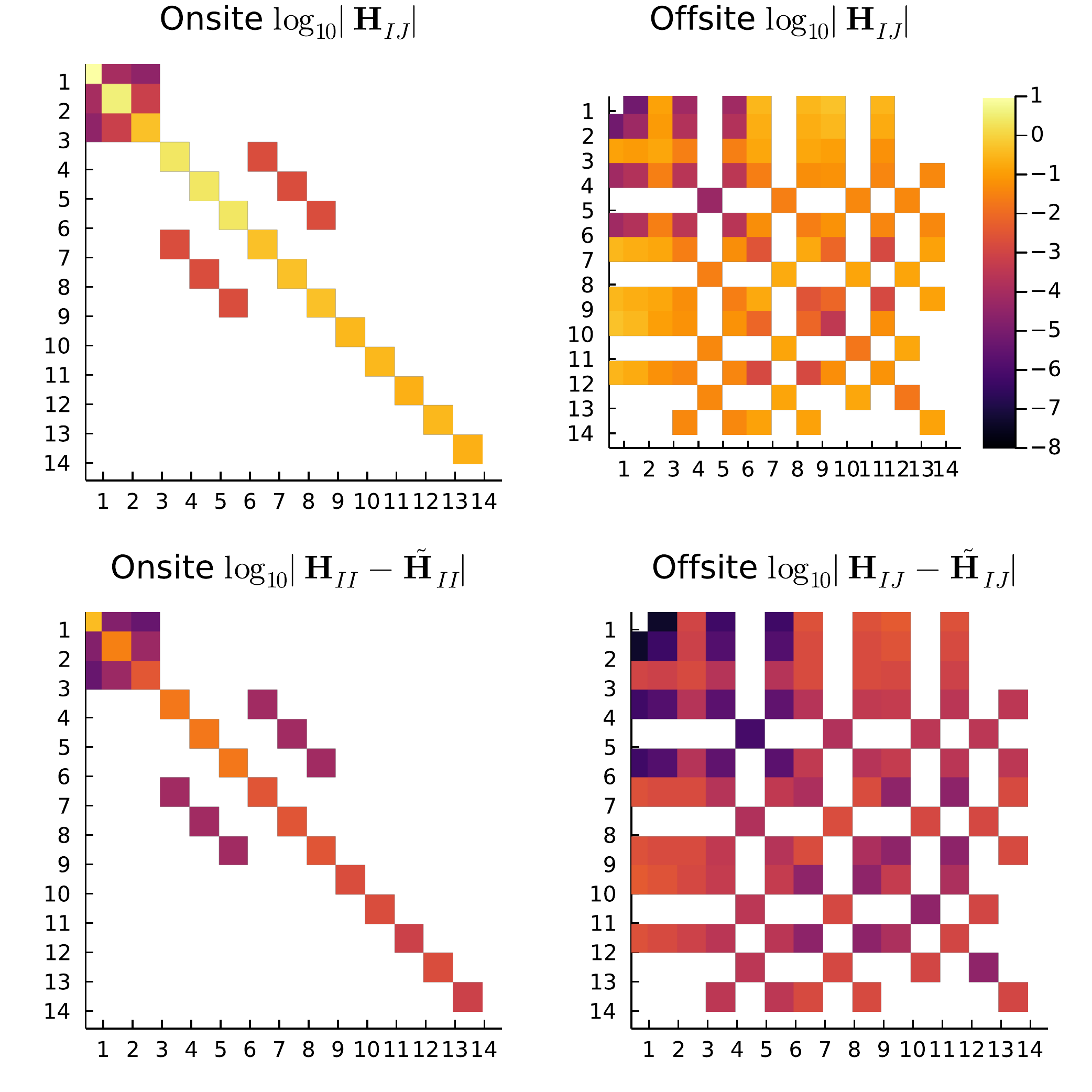}
\caption{Magnitudes (above) and errors (below) for onsite and offsite $\tilde{\bH}$ for prediction on the FCC ground state unitcell (not included in training set).}
    \label{fig:FCC_HS_errors}
\end{figure}

\subsection{Prediction of band structures and DoS}

So far we have assessed only errors made on the quantities used in  fitting the models, i.e. the Hamiltonian and overlap matrix elements. While it is reassuring that these are accurately captured, a stronger test of the predictive power of our formulation is to use it to predict electronic observables such as the band structure and DoS.
Fig.~5 compares predictions of these quantities for FCC and BCC aluminium  with those computed from the reference FHI-aims Hamiltonian and overlap matrices. There is excellent agreement for all occupied bands, and also bands within 10 eV of the Fermi level (which is itself in close agreement between the reference and predicted systems).
The DoS was integrated on a dense $9\times9\times9$ $k$-point mesh and also shows excellent agreement for the occupied states for both FCC and BCC, with significant errors only arising well above the Fermi level, giving confidence in the ability of our model to be predict electronic observables.

The figure also shows confidence intervals for the predicted band structures. These have been estimated to leading order using a simple \emph{a priori} error analysis to propagate errors in the Hamiltonian $\Delta \bH = \tilde{\bH} - \bH$ and overlap $\Delta \bS = \tilde{\bS} - \bS$ to expected errors in the bands using the result~\cite{crandall1973bifurcation}
\begin{equation}
\tilde{\epsilon} - \epsilon \sim \langle \phi | \Delta \bH - \epsilon \Delta \bS | \phi \rangle
\end{equation}
in the limit as $\Delta H, \Delta S \to 0$, where $\phi$, $\epsilon$ and $\tilde{\phi}$, $\tilde{\epsilon}$ are eigenfunctions and eigenvalues of the reference and approximated systems, respectively. Repeating this for each $k$-point leads to the error bounds shown. The error estimates prove reliable: the DFT bands, shown in red, are almost always contained within the blue shaded region.

\begin{figure}
    \centering
    \includegraphics[width=\columnwidth]{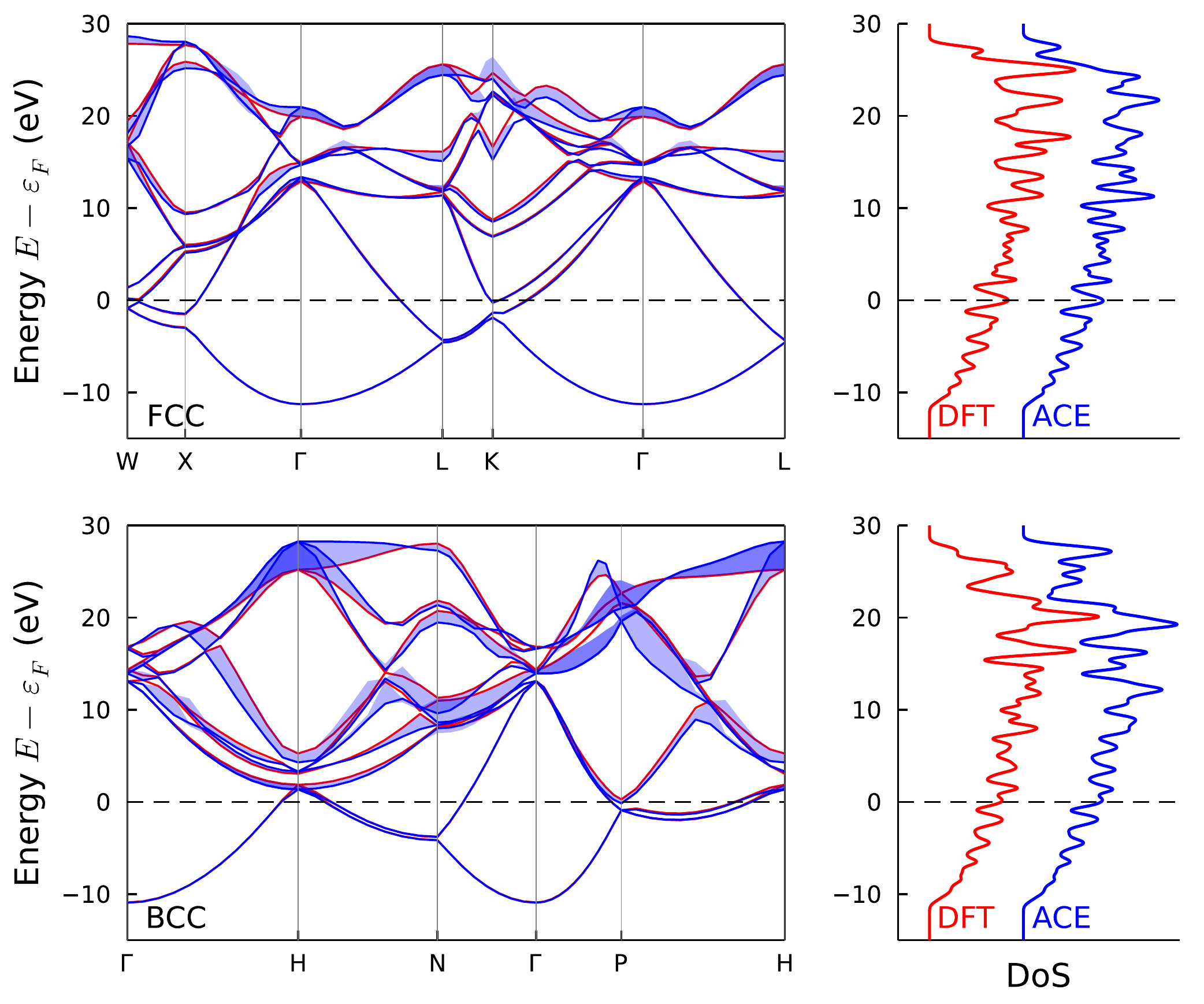}
    \caption{FCC and BCC band structures obtained with DFT (red) and predicted by an ACE model with onsite H order 2, and offsite H and S order 1 (blue). Confidence intervals shown with blue ribbons are from \emph{a priori} analysis of the errors in band spectrum expected to result from known errors in $\tilde{\bH}$ and $\tilde{\bS}$ (see text). \revision{Energies are shown relative to the DFT Fermi level.}}
    \label{fig:bands}
\end{figure}

Fig.~6 shows the convergence of bandstructures and DoS with respect to the maximum polynomial degree used in the ACE basis set, and for two choices of correlation order $\nu=1$ and $\nu=2$. 

The error in the DoS is computed using the first Wasserstein (or `earthmover') distance between the reference and predicted DoS,
\revision{ which is a natural metric for comparing densities of states since it is a distance between probability distributions (see, e.g., \cite{PhysRevB.102.235130}).}
The error in band structures is defined as the RMSE in the $k$-dependent band energies
\begin{equation}
    E_\mathrm{band}(\mathbf{k}) = \sum_{i=1}^{N_\mathrm{orb}} f\left(\frac{\epsilon_i - \varepsilon_F}{\sigma}\right) \epsilon_i(\mathbf{k}) \label{eq:band-error}
\end{equation}
along the high-symmetry $k$-paths shown in Fig.~5, 
where $f(\bullet)$ is the Fermi function,  $\varepsilon_F$ is the Fermi level of the system and the smearing width is taken to be $\sigma=0.086$~eV, corresponding to an electronic temperature of 1000~K.

The models with untuned parameters shown with the solid lines and dashed lines in Fig.~6 are already sufficiently accurate to produce good band structures and densities of states. However, when increasing the maximum degree used for all subblocks simultaneously, some overfitting can be seen, similar to that observed in the direct validation results of Fig.~3, and once again this arises at lower degrees of 9--12 with $\nu=2$ than with $\nu=1$, where maximum degrees of up to 13--14 are possible without overfitting. Errors in the DoS and the band structure for both FCC and BCC are further reduced when using the optimised model of \S~\ref{sec:optimal-model}, shown with the horizontal dotted lines in the figure to produce band structures with a RMSE of less than 0.4~eV for both phases.

\begin{figure}
    \centering
    \includegraphics[width=\columnwidth]{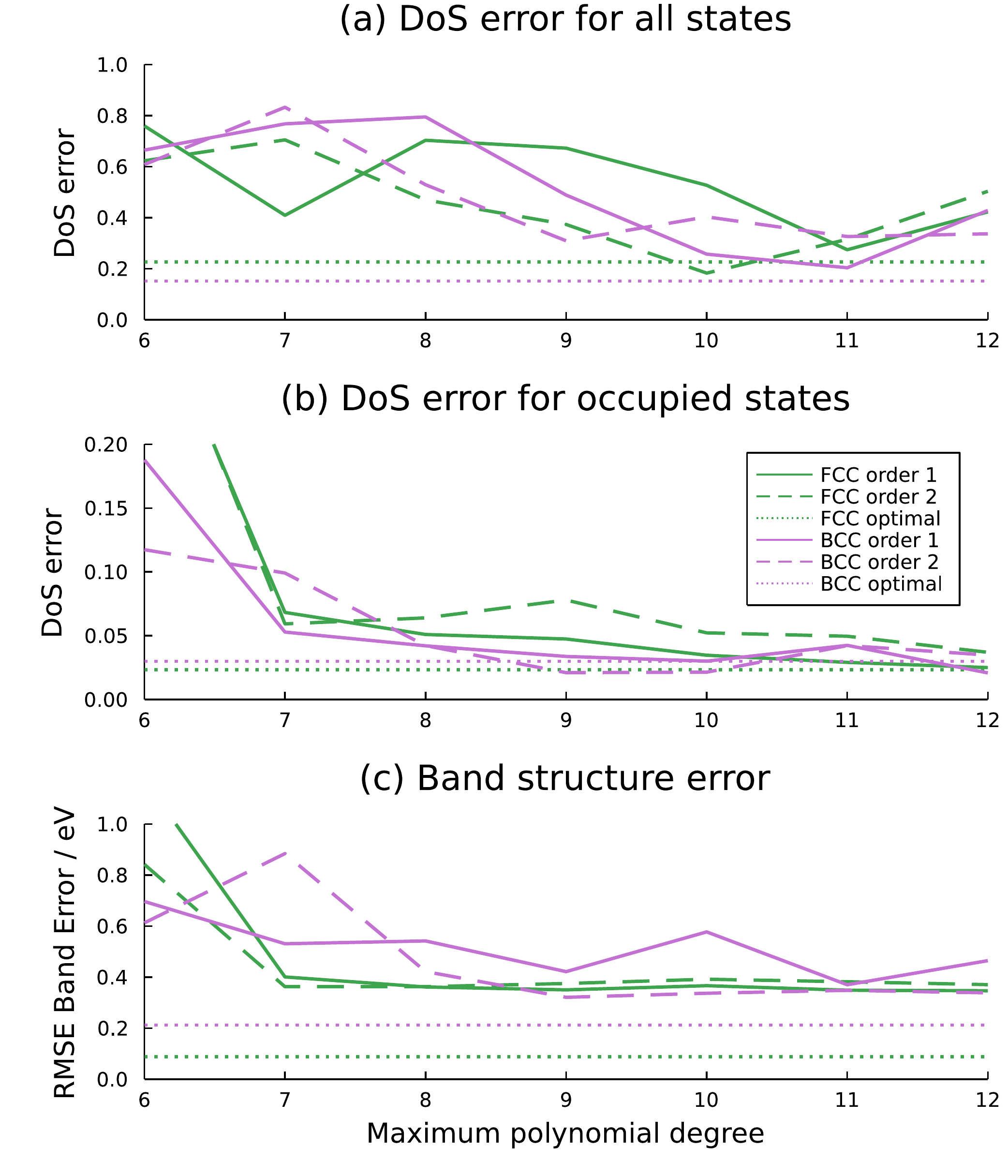}
    \caption{Convergence of FCC and BCC band structures and DoS with respect to the correlation order $\nu$ and maximum polynomial degree $d_\mathrm{max}$ used in the ACE basis set. Dotted lines show the optimised model of \S~\ref{sec:optimal-model}.
    \revision{ (a) Error in the full DoS. (b) Error in the occupied states, i.e. those below the Fermi level. (c) Band error computed with Eq.~\eqref{eq:band-error}.}
    }
    \label{fig:bands_DoS_convergence}
\end{figure}

\subsection{BCC to FCC transition}

As a challenging test, we used our optimised model to predict the Hamiltonian and overlap matrices along the Bain transformation path from BCC to FCC. We then diagonalised the predicted matrices to obtain the eigenvalues and hence the DoS at each point along the path and compared to reference values computed with FHI-aims for the same systems. As can be seen in Fig.~7, the predicted electronic structure agrees well at all points along the path, suggesting good extrapolative behaviour beyond the training set,  which includes only environments accessible from the two minima at moderate temperatures during MD.

Notably, nowhere along the path is the accuracy of the ACE model worse than it is for FCC or BCC, although accuracy drops off outside the BCC--FCC range $1/\sqrt{2} < c/a < 1$. 
We interpret this as meaning that along the Bain path we see different global structures, but similar local environments, whereas to the left of BCC and to the right of FCC we go outside the range of local environments included in the training set.

\begin{figure}
    \centering
    \includegraphics[width=\columnwidth]{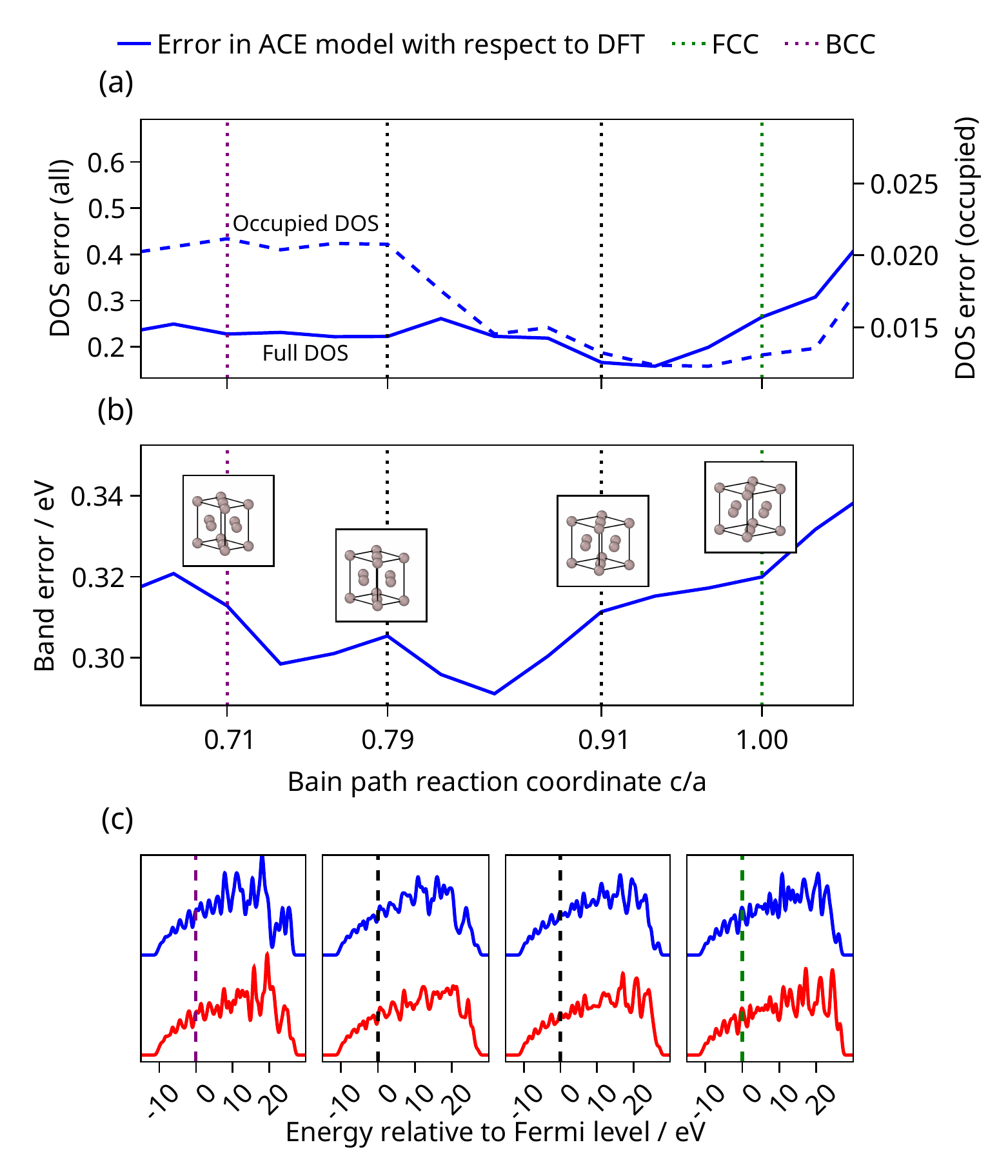}
    \caption{Electronic structure along the transition from BCC $c/a=1/\sqrt{2}\approx 0.71$ to FCC $c/a=1$. (a) Error in the density of states made by our ACE model with respect to the DFT reference (measured with the Wasserstein distance) as a function of $c/a$ along the Bain path. \revision{ The solid line shows the full error in the DoS (right vertical axis), while the dashed line shows the error in the occupied states (left vertical axis; note the change of scale).} (b) RMSE error in the electronic band structure (along high-symmetry $k$-path for the BCC structure) as a function of $c/a$ along the Bain path. Insets illustrate the structure of the cubic cell at points along the path. (c) Comparisons of densities of state for the ACE model (blue) and DFT (red) at four points along the path, including the BCC (left) and FCC (right) structures.}
    \label{fig:BCC_to_FCC}
\end{figure}

\subsection{\revision{Restricted training databases}}

To further test \revision{how well the model generalises across crystal systems}, we carried out two further fits using the same optimal parameters as for the final model presented above, but with the training database restricted to either FCC only or BCC only configurations (using subsets of the same MD-generated structures as above). We then checked the ability of the resulting ACE Hamiltonian models to predict the DFT electronic structure of both crystals. The results, illustrated in Figure.~8 and summarised in Table~\ref{tab:crossover} convincingly demonstrate the approach has excellent transferability, since the FCC DoS (and also the associated full band structure) can be accurately predicted using only BCC training data, and \emph{vice versa}.

\begin{figure}
    \centering
    \includegraphics[width=\columnwidth]{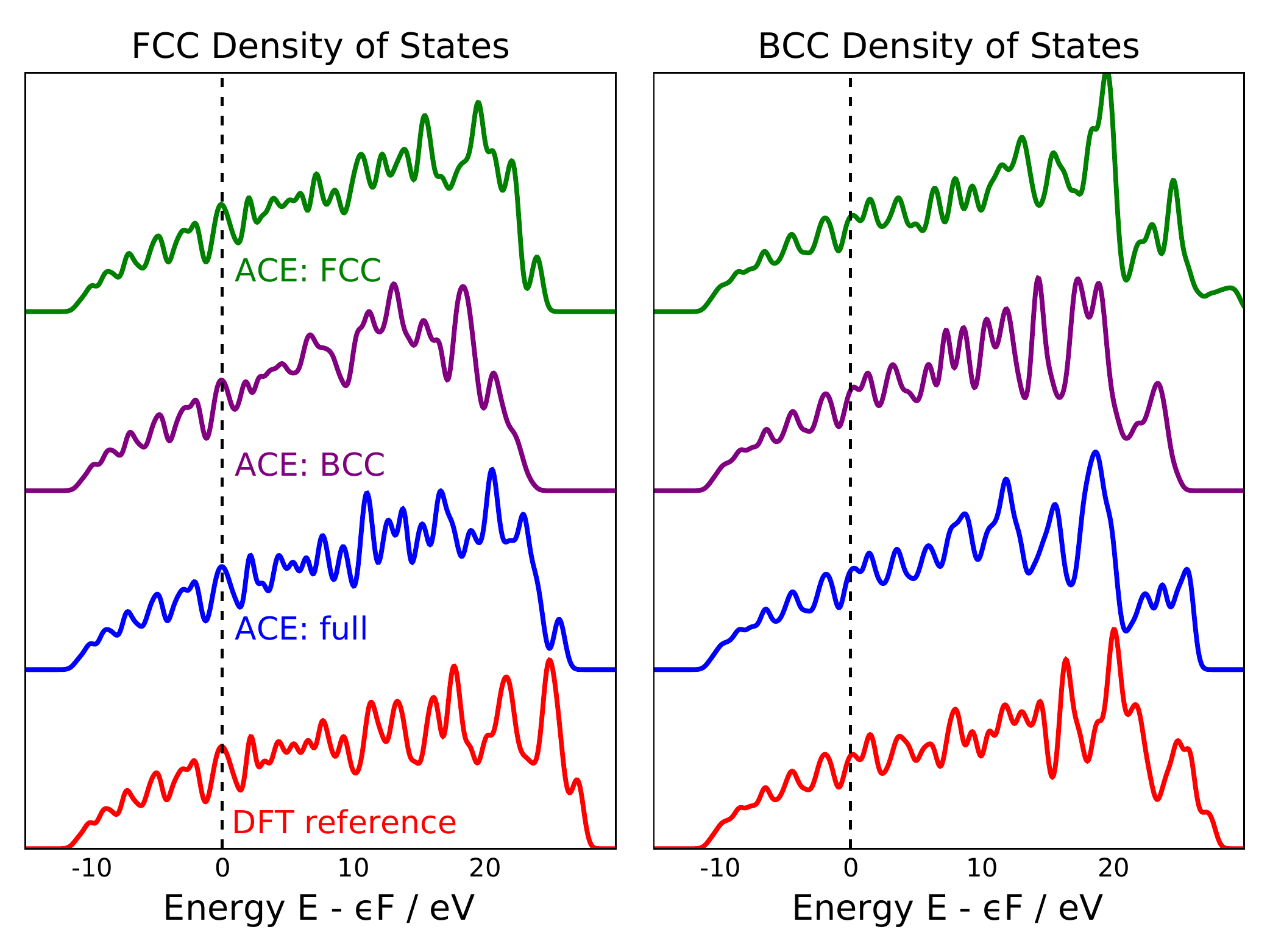}
    \caption{\revision{Comparison of FCC and BCC DoS predicted with ACE models for full and restricted training databases. The reference DFT DoS is shown in red. A vertical shift has been applied to separate the DoS for each ACE model. }}
    \label{fig:crossover}
\end{figure}

\begin{table}[]
    \centering
    \begin{tabular}{c|c|c|c}
      Crystal & Training database & DoS error (all) & DoS error (occ.) \\
\hline      
FCC & FCC+BCC & 0.424 & 0.015 \\ 
FCC & BCC & 0.930 & 0.081 \\ 
FCC & FCC & 0.732 & 0.044 \\ 
\hline
BCC & FCC+BCC & 0.308 & 0.023 \\ 
BCC & BCC & 0.550 & 0.041 \\ 
BCC & FCC & 0.311 & 0.025 \\          
    \end{tabular}
    \caption{\revision{ Errors in the FCC and BCC DoS predicted with ACE models for full and restricted training databases. The error in the full DoS and in the occupied states below the Fermi level are reported.}}
    \label{tab:crossover}
\end{table}

\subsection{\revision{Defected structures}}

\revision{As a final test of our models' ability to predict outside of the domain of the training sets, we predicted the electronic structure of a 728 atom $9 \times 9 \times 9$ FCC Aluminium supercell containing a single vacancy. The structure was obtained by deleting an atom from the supercell and performing a geometry optimisation with FHI-aims until the maximum force was less than $5\times10^{-3}$~eV/\AA{}. We then compared the projected DoS (PDoS) for the atomic orbitals neighbouring the vacancy as obtained with DFT with the predictions of our optimal ACE Hamiltonian model, without refitting.
The DFT and ACE PDoS are shown in Fig.~9 and demonstrate convincingly that our model is able to capture the changes in the local electronic structure associated with the introduction of a defect, indicating that it correctly predicts the self consistent field (SCF) relaxations of the Hamiltonian without a need for an explicit SCF loop in the approximate scheme.}

\begin{figure}
    \centering
    \includegraphics[width=\columnwidth]{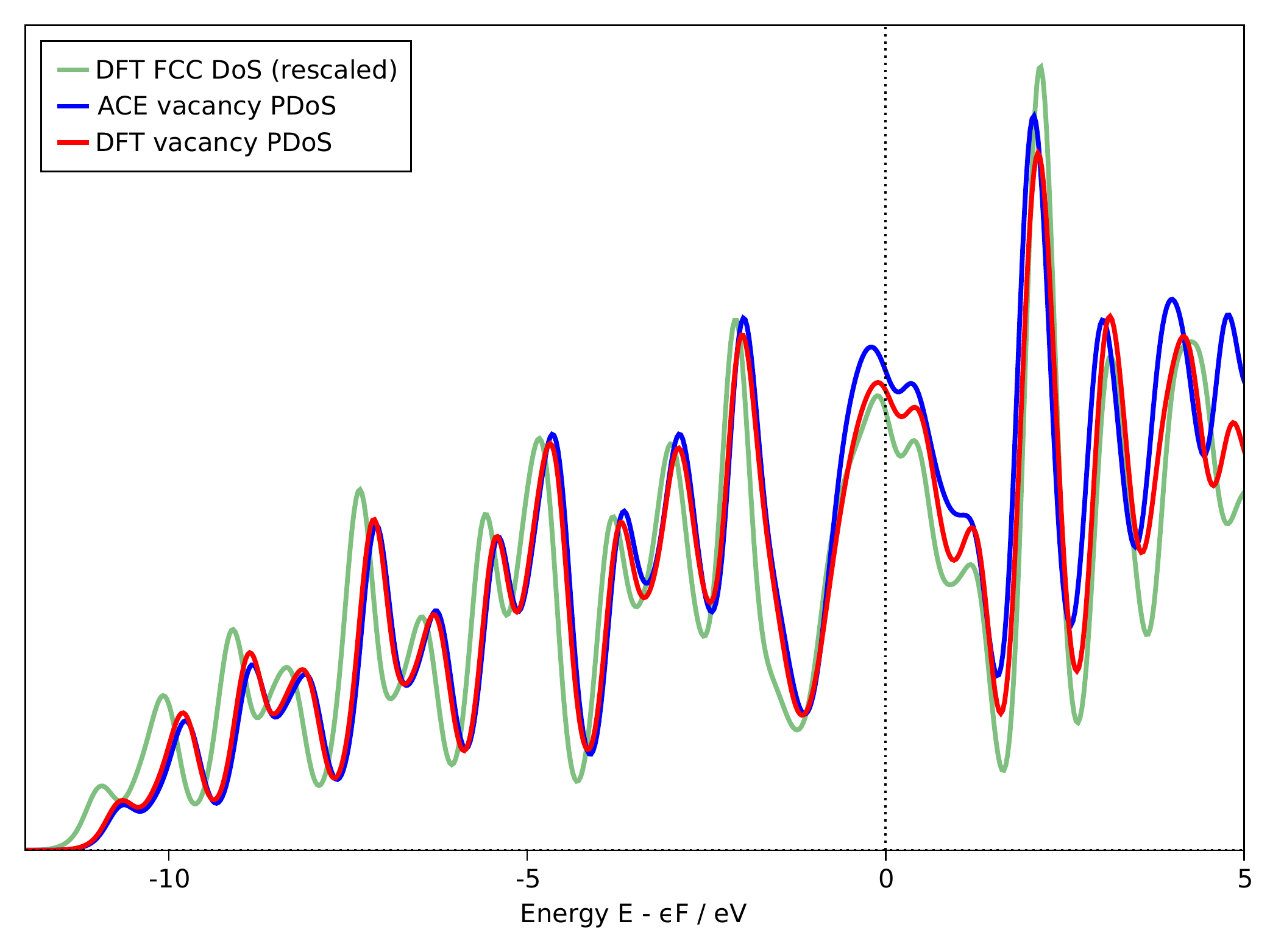}
    \caption{\revision{Comparison of  PDoS for a 728-atom Al vacancy supercell between the reference DFT results (red) and those predicted by our ACE Hamiltonian model (blue), which was not trained on vacancy data. PDoS includes orbitals associated with the nearest neighbours of the vacancy. The PDoS for the perfect FCC structure is shown in green to allow the changes in the electronic structure due to the defect to be assessed.}}
    \label{fig:vacancy}
\end{figure}

%}

\section{Discussion} \label{sec:conclusion}

We have reported a data-driven scheme to construct predictive models of Hamiltonian and overlap matrices from \emph{ab initio} data. Our scheme incorporates all relevant symmetry operations, giving an equivariant analytical map from first principles data to linear models for the Hamiltonian and overlap matrices as a function of the atomic and bond environments. We have shown that it is possible to apply our methodology to produce accurate predictions for the band structure in aluminium in both FCC and BCC phases from limited training data. The approach has huge potential for delivering comparable accuracy to DFT while at the same time reaching time and length scales far beyond its capabilities. For example, it opens the door to high-throughput computation of quantities which depend on electronic properties, such as photoemission spectra, transport coefficients, and electron-phonon coupling constants, all of which can currently only be accurately computed with first principles methods~\cite{Knoop2020}.

Our results are extremely encouraging, and there are a number of avenues open for further exploration. From a computational performance perspective, we note the evaluation of Hamiltonian and overlap blocks is trivially parallelisable with perfect scaling. Performance enhancements would also come from further optimisation of the ACE basis used to represent the Hamiltonian and overlap matrices, e.g. by sparsifying to reduce the basis set size, or by incorporating non-linearity to reduce the maximum degree required~\cite{Lysogorskiy2021}. Moreover, Bayesian approaches to model selection could be used instead of cross-validation. This would lead to more efficient model construction, as well as the possibility of \emph{a priori} error estimates on the accuracy of model predictions through  uncertainty propagation.

Further comprehensive studies of the dependence of accuracy and transferability of models on quantity and type of training data, as well as extension to materials and system with more complex bonding environments are also necessary. In future we will expand this approach to explore multi-component systems. A further extension will be to fit a potential $\tilde{E}$ to allow total energy and forces to be predicted by adding a correction to the band energy. For example, $\tilde{E}$ could be represented by an ACE potential determined from the local atomic environments.

\section{Methods} \label{sec:methods}

\subsection{Data generation}

The datasets used in this work are constructed for face-centred cubic (FCC) and body-centred cubic (BCC) phases of Al. Our data was generated through electronic structure calculations with the all-electron numeric atomic orbital code FHI-aims (version 190530)~\cite{Blum2009}. We used the Perdew-Burke-Enzerhof (PBE) generalized gradient approximation \cite{PBE} to the exchange-correlation energy within the KS-DFT formulation, and neglected spin in our treatment.  The convergence criteria for charge density, sum of eigenvalues, and total energy of the self-consistent cycles were set to $10^{-5}$~e/a$_0^3$, 5$\times10^{-5}$~eV, and $10^{-6}$~eV, respectively. 
%The Pulay density mixing algorithm was used with a maximum of 8 past iterations to be mixed in the SCF convergence. 
The default {\em tight} FHI-aims  basis set and integration grid  definitions were used, which use a basis set confinement with a maximum radial basis function extent of 6\,\AA. 
%set the confinement radius of each potential to start at 4\,\AA\ with a smooth cutoff at 6\,\AA\.
We modify the set of atomic basis functions that we employ to achieve optimal computational efficiency. Systematic convergence tests showed that band energies are converged up to 10 eV above the Fermi level when using a minimal basis plus a single {\em d} orbital from Tier 1. Therefore, we used a basis set comprising {\em s} and {\em p} orbitals of the minimal basis set plus one {\em d} orbital from  the Tier 1 setting, yielding the 14 atomic basis functions for Al illustrated in Fig.2(b).

%To determine the atomic basis set for electronic calculations, we tested various basis definitions starting from the minimal basis set of {\em s} and {\em p} orbitals. We tested adding {\em d} and {\em f} orbitals up to Tier 2 as parameterised in FHI-aims. Convergence tests showed that band energies are converged up to 10 eV above the Fermi level when using a minimal basis plus a single {\em d} orbital from Tier 1. Therefore, we used a basis set comprising {\em s} and {\em p} orbitals of the minimal basis set plus one {\em d} orbital from  the Tier 1 setting, yielding the 14 electron basis set illustrated in Fig.2(b).

%We carried out geometry optimizations to determine the equilibrium lattice constants for FCC and BCC Al. 
The optimal equilibrium lattice constants for FCC and BCC Al were determined in primitive cells with a $9\times9\times9$ Monkhorst-Pack $\mathbf{k}$-point mesh~\cite{Monkhorst1976} to be  4.05\,\AA\ and 3.29\,\AA\,  respectively.
To sample a variety of distorted atomic configurations for Al, we carried out molecular dynamics (MD) simulations at a temperature of 500~K using $9\times9\times9=729$ atom supercells of the primitive FCC and BCC unit cells. MD simuations for each phase were performed in the $NPT$ ensemble using a 5 fs timestep and the Embedded Atom Method (EAM) potential proposed by Zhou \emph{et al.}~\cite{zhou2004}. Single point DFT total energy calculations were carried out on the final configurations of each of these 500 MD simulations using FHI-aims with the parameters described above and a single  $\mathbf{k}$-point at $\Gamma$. We stored the resulting $H$ and $S$ matrices giving a dataset
\begin{align}
    & \big\{ (H_{II}, \xR_I) \big\}, \\ 
    & \big\{ (H_{IJ},  \bR_{IJ}, \xR_{IJ})  \big\}, \\ 
    & \big\{ (S_{IJ},  \bR_{IJ}, \xR_{IJ})  \big\}. 
\end{align}
\noindent where $II$, $IJ$ indicate on- and off-site blocks of the Hamiltonian and overlap matrices while $\bR_\bullet$ and $\xR_\bullet$ are the corresponding atomic structure data as defined in \ref{sec:2A}. \revision{ For the optimised model reported in \S\ref{sec:optimal-model}}, we used \revision{1000}
%\sout{730}
training and \revision{1000}
%\sout{728}
test blocks for the onsite part of the Hamiltonian and  2000 training and \revision{2000}
%\sout{1369}
test blocks for the offsite Hamiltonian and overlap matrices \revision{(with more offsite than onsite data to reflect the far greater number of offsite blocks in the target matrices)}. Equal numbers of samples were taken from the FCC and BCC MD data.

\subsection{Parameter Estimation}
\label{sec:parameter_choices}
We have defined three linear models for equivariant components of Hamiltonian and overlap matrices (up to the choice of approximation parameters). It remains to specify a parameter estimation procedure to determine the model parameters which typically number in the thousands to tens of thousands. There are essentially two choices we can make: (i) fit the models to observed properties such as band structure, energies, forces; or (ii) fit the models directly to match a reference Hamiltonian. Both approaches have advantages and disadvantages. We have chosen to follow route (ii) which is particularly attractive from both theoretical and numerical perspectives as it results in a linear least squares problem. 

% {\it Step 2. Least squares system: }
Let $\tilde{\bK} = {\bm c} \cdot \mathcal{B}$ be one of the three linear models, and $\{ (\bK_*^{(\tau)}, \xR_\bullet^{(\tau)} \}_{\tau}$ the corresponding training set, then we set up the loss function 
\begin{equation}
    L_0({\bm c}) 
    = 
    \sum_{\tau} \big| \bK_*^{(\tau)} - \tilde{\bK}(\xR_\bullet^{(\tau)}) \big|^2.
\end{equation}
Since $\tilde{\bK}$ is linear in ${\bm c}$ it follows that $L$ can be rewritten as 
\begin{equation}
    L_0({\bm c}) = \big\| \xB {\bm c} - {\bm y} \big\|^2, 
\end{equation}
where $\xB$ is the design matrix and ${\bm y}$ contains the reference model values. To prevent overfitting, we regularise the least squares system with a generalised Tychonov term, 
\begin{equation} \label{eq:reglsq}
    L_\lambda({\bm c}) := 
        \big\| \xB  {\bm c} - {\bm y} \big\|^2
        + \lambda \big\| \Gamma {\bm c} \big\|^2, 
\end{equation}
where $\Gamma = {\rm diag}(\Gamma_{kk})$ with $\Gamma_{kk}$ an estimate for the curvature of the $k$th basis function which enforces smoothness of the model~\cite{Dusson2019-gn,Lysogorskiy2021} and $\lambda$ is a regularisation parameter. Throughout this work, we define $\Gamma_{kk}$ by
\begin{equation}
    \Gamma_{kk} = \sum_\nu (n_\nu^2+l_\nu^2+m_\nu^2), 
\end{equation}
and $\lambda$ is always set to be $10^{-7}$.
% \cco{maybe we should specify this precisely? Maybe in the supplement? TODO Liwei please }
We then solve the regularised least squares system \eqref{eq:reglsq} using an iterative LSQR algorithm with termination tolerance $10^{-6}$.

For the radial basis set $P_{nl}$ we used
    \begin{align}
        \xi(r) &= \Big(\frac{1 + r_0}{1 + r}\Big)^2\\ 
        P_{nl}(r) &= Q_n( \xi(r) )
    \end{align}
where $Q_n$ is a polynomial of degree $n$ such that 
$\int_{\xi_0}^{\xi_1} Q_n(\xi) Q_{n'}(\xi) d\xi = \delta_{nn'}$ and $[\xi_0, \xi_1] = \xi([0, r_{\rm cut}])$; see Ref.~\cite{Dusson2019-gn} for full details.

The envelope function for both on-site term and off-site environment basis function is defined as 
\begin{equation}
\fcut(r;r_{\rm cut}) = \fcutb(r;r_{\rm cut}) = \begin{cases}(r^2/r_{\rm cut}^2 - 1)^2,& r\leq r_{\rm cut},\\0, &r>r_{\rm cut},\end{cases}
\end{equation}
and that for offsite environment is given by a bond-related cylindrical cutoff function 
\begin{eqnarray*}
        && \fcute(z,r;z_{\rm cut},r_{\rm cut}) \\
       =&& \begin{cases}\Big(\frac{r^2}{r_{\rm cut}^2} - 1\Big)^2&\Big( \frac{z^2}{(z_{\rm cut}+l_{\rm bond}/2)^2} - 1\Big)^2, \\
        & r\leq r_{\rm cut},|z|\leq z_{\rm cut}+l_{\rm bond}/2,\\
       0, &{\rm otherwise},\end{cases}
\end{eqnarray*}
where $(z,r,\theta)$ are the cylindrical coordinates of an environment atom (though $\theta$ is not used in this definition) and $l_{\rm bond}$ is the length of the corresponding bond. Note that both $\fcut$ and $\fcutb$ are rotation invariant, they will not influence the equivariance of the basis at all. Meanwhile, though the cylindrical curoff function $\fcute$ is bond-dependent, it can be easily checked that it does no harm to rotation symmetry as well.
%\cco{possibly add a proof of this in the appendix/suppl}

In our implementation, the on-site cutoff $r_{\rm cut}$ is chosen to be $9.0$~\AA{} for $\bH_{\rm on}$ and the off-site bond cutoff is set to be $10.0$~\AA{}. %\clz{Could add a figure to explain this choice, is it necessary?}.
We set $r^{\rm e}_{\rm cut} = z^{\rm e}_{\rm cut} = 5.0$~\AA{} for the off-site environment.
   
%    \lz{unit is missing which I am not pretty sure...}
As noted above, we used correlation order $\nu=0$ for the offsite overlap $\bS_{II}$ since these blocks are not environment dependent. For $\bH_{II}$ we used correlation order $\nu=2$ throughout, while for $\bH_{IJ}$ we tested correlation orders of both $\nu=1$ and $\nu=2$. The maximum polynomial degree was chosen on a case-by-case basis to control the balance between accuracy and transferability through a cross-validation procedure as discussed in more detail in \S~\ref{sec:results} in the main text.

\subsection{Prediction}

The software implementation of our method follows the workflow illustrated in Figure 1. The Julia packages {\tt ACE.jl}~\cite{ACE_jl} and {\tt ACEhamiltonians.jl} implement the general Atomic Cluster Expansion basis sets and the specialisation to fitting and predicting Hamiltonians, respectively. 
%We predict the real space matrices $\bH(\xR)$ and $\bS(\xR)$.
%\clz{the final sentence seems incomplete? Did any of us try to mention something here or just a missing period?}
Given an input configuration $\xR$ we use the scheme described above to predict $\tilde{\bH}_\mathrm{on}(\xR), \tilde{\bH}_{\text{off}}(\xR)$ and $\tilde{\bS}_{\text{off}}(\xR)$. We then assemble complete approximate Cartesian Hamiltonian and overlap matrices $\tilde{\bH}$ and $\tilde{\bS}$ from the predicted blocks.  We can construct  $\mathbf{k}$-dependent variants and associated bandstructures via a standalone Julia implementation contained within the {\tt ACEhamiltonians.jl} package.

Using either the reference or the predicted matrices we can solve the generalised eigenproblems of the form
\begin{eqnarray}
    \bH(\mathbf{k}) \phi_i & = & \epsilon_i \bS(\mathbf{k}) 
    \phi_i \\
    \tilde{\bH}(\mathbf{k}) \tilde{\phi}_i & = & \tilde{\epsilon}_i \tilde{\bS}(\mathbf{k}) \tilde{\phi}_i 
\end{eqnarray}
to obtain $k$-dependent band energies $\epsilon_i, \tilde{\epsilon_i}$ and orbitals (eigenfunctions) $\phi_i, \tilde{\phi}_i$ for the reference and predicted systems respectively, where $i=1,\ldots, N_\mathrm{orb}$ and in this work $N_\mathrm{orb}=14$. Band structures, density of states (DoS) and other derived quantities can be computed by post-processing the band energies following standard practices.

\subsection{\label{sec:appendix_bloch} Transformation of H and S from real to reciprocal space representation}

According to Bloch's theorem, in crystal-periodic structures, the Hamiltonian and overlap matrices defined in terms of real-space atomic orbitals can be transformed into a block-diagonal form and solved via a set of $N_k$ independent generalised eigenvalue problems where each block corresponds to a vector $\bk$ within the reciprocal unit cell:
\begin{equation}
  \mathbf{H}(\bk) \psi_{i\bk} = \epsilon_{i\bk}\mathbf{S}(\bk)\psi_{i\bk} \quad i = 1,2,\cdots
\end{equation}
where $\psi_{\nu\bk}$ are Bloch wave functions and $\mathbf{H}(\bk)$ and  $\mathbf{S}(\bk)$ are Hamiltonian and overlap matrices defined in terms of a discrete crystal-periodic basis. 

For this, we define crystal-periodic generalised basis functions $\chi_{a,\bk}$ from real-space basis functions as follows:
\begin{equation}\label{eq:bloch-orbital}
    \chi_{a\bk}(\br) = \sum_N \exp{ \{ i\bk\cdot \mathbf{N}\mathbf{L} \} } \chi_a(\br + \mathbf{N}\mathbf{L}).
\end{equation}
%\clz{To keep this representation consistent to \eqref{eq: at_basis}, I'd suggest remove ``$-\bR_I$" here.}
%\cjrk{OK, done, and also changed to $\br$ instead of $\mathbf{r}$.}
In \eqref{eq:bloch-orbital}, $\mathbf{L}$ refers to the column matrix of lattice vectors and $\mathbf{N} = (N_1, N_2, N_3)$ is an index vector that specifies the position of the unit cell (in multiples of the lattice vectors) in which orbital $\chi_a$ is located.

The matrix elements of $ \mathbf{H}(\bk)$ and $ \mathbf{S}(\bk)$, respectively, are constructed via
\begin{align} \label{eq:k-dependent-H}
    &H_{ab} (\bk) = \braket{ \chi_{a\bk} | \hat{H} | \chi_{b\bk} } = \\ & \sum_{N,N'} \exp{ \{ i\bk\cdot (\mathbf{N}'-\mathbf{N})\cdot\mathbf{L} \} } \underbrace{\braket{ \chi_{a,N'} | \hat{H} | \chi_{b,N} }}_{=H_{ab}(\mathbf{N},\mathbf{N}')}
\end{align}
and
\begin{align} \label{eq:k-dependent-S}
    & S_{ab} (\bk) = \braket{ \chi_{a\bk} | \chi_{b\bk} } = \\ & \sum_{N,N'} \exp{ \{ i\bk\cdot (\mathbf{N}'-\mathbf{N})\cdot\mathbf{L} \} } \underbrace{\braket{ \chi_{a,N'} | \chi_{b,N} }}_{=S_{ab}(\mathbf{N},\mathbf{N}')}
\end{align}
where $H_{ab}(\mathbf{N},\mathbf{N}')$ and $S_{ab}(\mathbf{N},\mathbf{N}')$ are as defined in \eqref{eq:H} and \eqref{eq:S} for atomic orbitals defined in different unit cells $\mathbf{N}$ and $\mathbf{N}'$. 

In this work, we use this transformation to map the real-space matrices to arbitrarily dense $\bk$-grids as is common practice for localised basis sets such as atomic orbitals or maximally localized Wannier functions. We then calculate eigenvalues $\epsilon_{\nu\bk}$ at arbitrary points in reciprocal space to calculate converged electronic densities-of-state and band structures.

%As the $\bk$-dependent matrices and the solution of the set of generalised eigenvalues completely follow from the real-space $\bH$ and $\bS$, we will go on to develop a representation for those two matix quantities as a function of $\xR$.

\subsection{\label{sec:appendix_equivariant} Equivariance of $\bH_{IJ}$}
For the real space Hamiltonian $\mathbf{H}({\xR})$, we decompose it as $\mathbf{H}({\xR}) = \big(\bH_{IJ}\big)_{I,J=1}^{N_{\rm atom}}$ (cf. Fig.~2). Denote $a = (n,l,m; I) := (\alpha;I), b = (n',l',m'; J) := (\beta;J)$, we may then write
\begin{eqnarray*}
    \bH_{IJ}^{\alpha\beta}(\xR)= \langle\chi_a|\hat{H}|\chi_b\rangle.
\end{eqnarray*}

In the definition of $\chi_a$, the radial basis $R_{nl}(r)$ is invariant under rotation and $Y_{lm}(Q(\theta,\phi))$ can be expressed as linear combination of $Y_{l\mu}(\theta,\phi)$, i.e.,
\begin{equation}\label{eq:chi_equiv}
\chi_{(n,l,m;I)}(Q\br;Q\xR) = \sum_{\mu}D_{\mu m}^{l}\chi_{(n,l,\mu;I)}(\br;\xR).
\end{equation}
Here, $\chi_\bullet$ is $\xR$-dependent since it is atom-centred. Besides, 
\begin{eqnarray}\label{eq:hijab}
   &&\bH_{IJ}^{\alpha\beta}(Q\xR) \notag\\
   &=& \int_{\mathbb{R}^3} \chi_{a}(\br;Q\xR)^* V_{\text{eff}}(\br, Q\xR)\chi_b(\br;Q\xR)d\br \notag\\
   &=& \int_{\mathbb{R}^3} \chi_{a}(Q\br;Q\xR)^* V_{\text{eff}}(Q\br, Q\xR)\chi_b(Q\br;Q\xR)d\br \notag\\
   &=& \int_{\mathbb{R}^3} \chi_{a}(Q\br;Q\xR)^* V_{\text{eff}}(\br, \xR)\chi_b(Q\br;Q\xR)d\br.
\end{eqnarray}
Combining \eqref{eq:chi_equiv} and \eqref{eq:hijab}, we see immediately that  
\begin{equation}
   \bH_{IJ}(Q\xR) = D(Q)^* \bH_{IJ}(\xR) D(Q),
\end{equation}
where
\begin{equation}
    D(Q) = \textup{Diag}(D^{l_1}(Q), D^{l_2}(Q), \cdots),
\end{equation} 
and $D^{l_i}$ indicate the Wigner-D matrices.

Sometimes, the angular term in \eqref{eq: at_basis} is chosen to use real spherical harmonics rather than complex ones, i.e.,
\begin{align}
	\chi_a (\br) &= R_{nl}(r) S_{lm}(\theta, \phi), \qquad \text{and}  \\
    S_{lm} &= \sum_{m'}C_{mm'}Y_{lm},
\end{align}
where $\{C_{mm'}\}$ are the corresponding transforming coefficients. Equivalently, we may go through all possible indices $m$ with respect to a fixed $l$ and obtain the following matrix form
\begin{equation}
    {\bf S}_l(\xR) = C_l{\bf Y}_l(\xR).
\end{equation}
In this case, the equivariance of $\bH_{IJ}$ simply follows, just with the varied equivariant matrix
\begin{equation}
    \tilde{D}(Q) = \textup{Diag}(\tilde{D}^{l_1}(Q), \tilde{D}^{l_2}(Q), \cdots),
\end{equation} 
and $\tilde{D}^{l_i}(Q) = C_{l_i}D^{l_i}(Q)$.

\section{Data Availability}

Supporting data for this manuscript comprising the electronic structure training data, ACE Hamiltonian and overlap models, prediction results and an archived copy of the source code is available from \revision{\url{https://dx.doi.org/10.5281/zenodo.6561452}}.

\section{Code Availability}

The \texttt{ACE.jl} package which implements the Atomic Cluster Expansion basis sets used here is available from \url{https://github.com/acesuit/ACE.jl}. The \texttt{ACEhamiltonians.jl} package which extends its capabilities to learning Hamiltonian and overlap \revision{ matrices is available from \url{https://github.com/ACEsuit/ACEhamiltonians.jl/tree/arXiv.2111.13736}; with examples provided at \url{https://github.com/ACEsuit/ACEhamiltoniansExamples}.}

%\null\newpage

\section{Acknowledgments}
This work was financially supported by a Leverhulme Trust Research Project Grant (RPG-2017-191), the Engineering and Physical Science Research Council (EPSRC) under grant EP/R043612/1, the NOMAD Centre of Excellence (European Commission grant agreement ID 951786), and the UKRI Future Leaders Fellowship programme (MR/S016023/1). We acknowledge computational resources provided by the Scientific Computing Research Technology Platform of the University of Warwick, the EPSRC-funded HPC Midlands+ consortium (EP/P020232/1, EP/T022108/1) and on ARCHER2 (https://www.archer2.ac.uk/) via the UK Car-Parinello consortium (EP/P022065/1).
For the purpose of Open Access, the author has applied a CC-BY public copyright licence to any Author Accepted Manuscript (AAM) version arising from this submission.

\section{Author Contributions}

RJM, CO and JRK designed the research. GA and BO generated the training data. BO \revision{ and AM} developed the workflow infrastructure linking from DFT calculations to parametrised Hamiltonians. LZ, GD, and CO designed and implemented the equivariant basis sets. LZ and JRK generated and validated the models for aluminium with input from all authors. All authors analysed the results, contributed to the manuscript and approved its final version.

\section{Competing Interests}
The authors declare no competing interests.

\providecommand{\noopsort}[1]{}\providecommand{\singleletter}[1]{#1}%


\begin{thebibliography}{10}
\expandafter\ifx\csname url\endcsname\relax
  \def\url#1{\texttt{#1}}\fi
\expandafter\ifx\csname urlprefix\endcsname\relax\def\urlprefix{URL }\fi
\providecommand{\bibinfo}[2]{#2}
\providecommand{\eprint}[2][]{\url{#2}}

\bibitem{Bitzek2015}
\bibinfo{author}{Bitzek, E.}, \bibinfo{author}{Kermode, J.~R.} \&
  \bibinfo{author}{Gumbsch, P.}
\newblock \bibinfo{title}{Atomistic aspects of fracture}.
\newblock \emph{\bibinfo{journal}{Int. J. Fract.}}
  \textbf{\bibinfo{volume}{191}}, \bibinfo{pages}{13--30}
  (\bibinfo{year}{2015}).

\bibitem{jiangDynamicsReactionsMetal2019}
\bibinfo{author}{Jiang, B.} \& \bibinfo{author}{Guo, H.}
\newblock \bibinfo{title}{Dynamics in reactions on metal surfaces: A
  theoretical perspective}.
\newblock \emph{\bibinfo{journal}{J. Chem. Phys.}}
  \textbf{\bibinfo{volume}{150}}, \bibinfo{pages}{180901}
  (\bibinfo{year}{2019}).

\bibitem{behlerFourGenerationsHighDimensional2021}
\bibinfo{author}{Behler, J.}
\newblock \bibinfo{title}{Four generations of high-dimensional neural network
  potentials}.
\newblock \emph{\bibinfo{journal}{Chem. Rev.}} \textbf{\bibinfo{volume}{121}},
  \bibinfo{pages}{10037--10072} (\bibinfo{year}{2021}).

\bibitem{unkeMachineLearningForce2021}
\bibinfo{author}{Unke, O.~T.} \emph{et~al.}
\newblock \bibinfo{title}{Machine learning force fields}.
\newblock \emph{\bibinfo{journal}{Chem. Rev.}} \textbf{\bibinfo{volume}{121}},
  \bibinfo{pages}{10142--10186} (\bibinfo{year}{2021}).

\bibitem{deringerGaussianProcessRegression2021}
\bibinfo{author}{Deringer, V.~L.} \emph{et~al.}
\newblock \bibinfo{title}{Gaussian process regression for materials and
  molecules}.
\newblock \emph{\bibinfo{journal}{Chem. Rev.}} \textbf{\bibinfo{volume}{121}},
  \bibinfo{pages}{10073--10141} (\bibinfo{year}{2021}).

\bibitem{musilPhysicsInspiredStructuralRepresentations2021}
\bibinfo{author}{Musil, F.} \emph{et~al.}
\newblock \bibinfo{title}{Physics-inspired structural representations for
  molecules and materials}.
\newblock \emph{\bibinfo{journal}{Chem. Rev.}} \textbf{\bibinfo{volume}{121}},
  \bibinfo{pages}{9759--9815} (\bibinfo{year}{2021}).

\bibitem{Mishin2021}
\bibinfo{author}{Mishin, Y.}
\newblock \bibinfo{title}{Machine-learning interatomic potentials for materials
  science}.
\newblock \emph{\bibinfo{journal}{Acta Mater.}} \textbf{\bibinfo{volume}{214}},
  \bibinfo{pages}{116980} (\bibinfo{year}{2021}).

\bibitem{Behler2021}
\bibinfo{author}{Behler, J.} \& \bibinfo{author}{Cs{\'a}nyi, G.}
\newblock \bibinfo{title}{Machine learning potentials for extended systems: a
  perspective}.
\newblock \emph{\bibinfo{journal}{Eur. Phys. J. B}}
  \textbf{\bibinfo{volume}{94}}, \bibinfo{pages}{142} (\bibinfo{year}{2021}).

\bibitem{Dewar1985}
\bibinfo{author}{Dewar, M.~J.}, \bibinfo{author}{Zoebisch, E.~G.},
  \bibinfo{author}{Healy, E.~F.} \& \bibinfo{author}{Stewart, J.~J.}
\newblock \bibinfo{title}{{AM}1: A new general purpose quantum mechanical
  molecular model1}.
\newblock \emph{\bibinfo{journal}{J. Am. Chem. Soc.}}
  \textbf{\bibinfo{volume}{107}}, \bibinfo{pages}{3902--3909}
  (\bibinfo{year}{1985}).

\bibitem{stewartOptimizationParametersSemiempirical1989}
\bibinfo{author}{Stewart, J. J.~P.}
\newblock \bibinfo{title}{Optimization of parameters for semiempirical methods
  i. method}.
\newblock \emph{\bibinfo{journal}{J. Comput. Chem.}}
  \textbf{\bibinfo{volume}{10}}, \bibinfo{pages}{209--220}
  (\bibinfo{year}{1989}).

\bibitem{Porezag95}
\bibinfo{author}{Porezag, D.}, \bibinfo{author}{Frauenheim, T.},
  \bibinfo{author}{Köhler, T.}, \bibinfo{author}{Seifert, G.} \&
  \bibinfo{author}{Kaschner, R.}
\newblock \bibinfo{title}{Construction of tight-binding-like potentials on the
  basis of density-functional theory: Application to carbon}.
\newblock \emph{\bibinfo{journal}{Phys. Rev. B}} \textbf{\bibinfo{volume}{51}},
  \bibinfo{pages}{12947} (\bibinfo{year}{1995}).

\bibitem{Elstner98}
\bibinfo{author}{Elstner, M.} \emph{et~al.}
\newblock \bibinfo{title}{Self-consistent-charge density-functional
  tight-binding method for simulations of complex materials properties}.
\newblock \emph{\bibinfo{journal}{Phys. Rev. B}} \textbf{\bibinfo{volume}{58}},
  \bibinfo{pages}{7260--7268} (\bibinfo{year}{1998}).

\bibitem{Sankey1989}
\bibinfo{author}{Sankey, O.~F.} \& \bibinfo{author}{Niklewski, D.~J.}
\newblock \bibinfo{title}{Ab initio multicenter tight-binding model for
  molecular-dynamics simulations and other applications in covalent systems}.
\newblock \emph{\bibinfo{journal}{Phys. Rev. B}} \textbf{\bibinfo{volume}{40}},
  \bibinfo{pages}{3979--3995} (\bibinfo{year}{1989}).

\bibitem{Lewis2001}
\bibinfo{author}{Lewis, J.~P.} \emph{et~al.}
\newblock \bibinfo{title}{Further developments in the local-orbital
  density-functional-theory tight-binding method}.
\newblock \emph{\bibinfo{journal}{Phys. Rev. B}} \textbf{\bibinfo{volume}{64}},
  \bibinfo{pages}{195103} (\bibinfo{year}{2001}).

\bibitem{bannwarthGFN2xTBAccurateBroadly2019}
\bibinfo{author}{Bannwarth, C.}, \bibinfo{author}{Ehlert, S.} \&
  \bibinfo{author}{Grimme, S.}
\newblock \bibinfo{title}{{GFN}2-{xTB}—an accurate and broadly parametrized
  self-consistent tight-binding quantum chemical method with multipole
  electrostatics and density-dependent dispersion contributions}.
\newblock \emph{\bibinfo{journal}{J. Chem. Theory Comput.}}
  \textbf{\bibinfo{volume}{15}}, \bibinfo{pages}{1652--1671}
  (\bibinfo{year}{2019}).

\bibitem{westermayrPerspectiveIntegratingMachine2021}
\bibinfo{author}{Westermayr, J.}, \bibinfo{author}{Gastegger, M.},
  \bibinfo{author}{Schütt, K.~T.} \& \bibinfo{author}{Maurer, R.~J.}
\newblock \bibinfo{title}{Perspective on integrating machine learning into
  computational chemistry and materials science}.
\newblock \emph{\bibinfo{journal}{J. Chem. Phys.}}
  \textbf{\bibinfo{volume}{154}}, \bibinfo{pages}{230903}
  (\bibinfo{year}{2021}).

\bibitem{Li2018}
\bibinfo{author}{Li, H.}, \bibinfo{author}{Collins, C.},
  \bibinfo{author}{Tanha, M.}, \bibinfo{author}{Gordon, G.~J.} \&
  \bibinfo{author}{Yaron, D.~J.}
\newblock \bibinfo{title}{A density functional tight binding layer for deep
  learning of chemical hamiltonians}.
\newblock \emph{\bibinfo{journal}{J. Chem. Theory Comput.}}
  \textbf{\bibinfo{volume}{14}}, \bibinfo{pages}{5764--5776}
  (\bibinfo{year}{2018}).
\newblock \eprint{1808.04526}.

\bibitem{stohrAccurateManyBodyRepulsive2020}
\bibinfo{author}{Stöhr, M.}, \bibinfo{author}{Medrano~Sandonas, L.} \&
  \bibinfo{author}{Tkatchenko, A.}
\newblock \bibinfo{title}{Accurate many-body repulsive potentials for
  density-functional tight binding from deep tensor neural networks}.
\newblock \emph{\bibinfo{journal}{J. Phys. Chem. Lett.}}
  \textbf{\bibinfo{volume}{11}}, \bibinfo{pages}{6835--6843}
  (\bibinfo{year}{2020}).

\bibitem{qiaoOrbNetDeepLearning2020}
\bibinfo{author}{Qiao, Z.}, \bibinfo{author}{Welborn, M.},
  \bibinfo{author}{Anandkumar, A.}, \bibinfo{author}{Manby, F.~R.} \&
  \bibinfo{author}{{MillerIII}, T.~F.}
\newblock \bibinfo{title}{{OrbNet}: Deep learning for quantum chemistry using
  symmetry-adapted atomic-orbital features}.
\newblock \emph{\bibinfo{journal}{J. Chem. Phys.}}
  \textbf{\bibinfo{volume}{153}}, \bibinfo{pages}{124111}
  (\bibinfo{year}{2020}).

\bibitem{supkaAFLOWpMinimalistApproach2017}
\bibinfo{author}{Supka, A.~R.} \emph{et~al.}
\newblock \bibinfo{title}{{AFLOW$\pi$}: A minimalist approach to
  high-throughput ab initio calculations including the generation of
  tight-binding hamiltonians}.
\newblock \emph{\bibinfo{journal}{Comp. Mater. Sci.}}
  \textbf{\bibinfo{volume}{136}}, \bibinfo{pages}{76--84}
  (\bibinfo{year}{2017}).

\bibitem{garrityDatabaseWannierTightbinding2021}
\bibinfo{author}{Garrity, K.~F.} \& \bibinfo{author}{Choudhary, K.}
\newblock \bibinfo{title}{Database of wannier tight-binding hamiltonians using
  high-throughput density functional theory}.
\newblock \emph{\bibinfo{journal}{Sci Data}} \textbf{\bibinfo{volume}{8}},
  \bibinfo{pages}{106} (\bibinfo{year}{2021}).

\bibitem{Marzari2012}
\bibinfo{author}{Marzari, N.}, \bibinfo{author}{Mostofi, A.~A.},
  \bibinfo{author}{Yates, J.~R.}, \bibinfo{author}{Souza, I.} \&
  \bibinfo{author}{Vanderbilt, D.}
\newblock \bibinfo{title}{Maximally localized wannier functions: Theory and
  applications}.
\newblock \emph{\bibinfo{journal}{Rev. Mod. Phys.}}
  \textbf{\bibinfo{volume}{84}}, \bibinfo{pages}{1419--1475}
  (\bibinfo{year}{2012}).
\newblock \eprint{1112.5411}.

\bibitem{Barzdajn2021}
\bibinfo{author}{Barzdajn, B.}, \bibinfo{author}{Garrett, A.~M.},
  \bibinfo{author}{Whiting, T.~M.} \& \bibinfo{author}{Race, C.~P.}
\newblock \bibinfo{title}{Development of data-driven spd tight-binding models
  of fe—parameterisation based on qsgw and dft calculations including
  information about higher-order elastic constants}.
\newblock \emph{\bibinfo{journal}{Model. Simul. Mater. Sci. Eng.}}
  \textbf{\bibinfo{volume}{29}}, \bibinfo{pages}{085006}
  (\bibinfo{year}{2021}).

\bibitem{Jenke2021}
\bibinfo{author}{Jenke, J.}, \bibinfo{author}{Ladines, A.~N.},
  \bibinfo{author}{Hammerschmidt, T.}, \bibinfo{author}{Pettifor, D.~G.} \&
  \bibinfo{author}{Drautz, R.}
\newblock \bibinfo{title}{Tight-binding bond parameters for dimers across the
  periodic table from density-functional theory}.
\newblock \emph{\bibinfo{journal}{Phys. Rev. Materials}}
  \textbf{\bibinfo{volume}{5}}, \bibinfo{pages}{023801} (\bibinfo{year}{2021}).

\bibitem{Blum2009}
\bibinfo{author}{Blum, V.} \emph{et~al.}
\newblock \bibinfo{title}{Ab initio molecular simulations with numeric
  atom-centered orbitals}.
\newblock \emph{\bibinfo{journal}{Comp. Phys. Commun.}}
  \textbf{\bibinfo{volume}{180}}, \bibinfo{pages}{2175--2196}.

\bibitem{Behler2011}
\bibinfo{author}{Behler, J.}
\newblock \bibinfo{title}{Atom-centered symmetry functions for constructing
  high-dimensional neural network potentials}.
\newblock \emph{\bibinfo{journal}{J. Chem. Phys.}}
  \textbf{\bibinfo{volume}{134}}, \bibinfo{pages}{074106}.
\newblock \bibinfo{note}{{ISBN}: 0021-9606}.

\bibitem{Bartok2013}
\bibinfo{author}{Bartók, A.~P.}, \bibinfo{author}{Kondor, R.} \&
  \bibinfo{author}{Csányi, G.}
\newblock \bibinfo{title}{On representing chemical environments}.
\newblock \emph{\bibinfo{journal}{Phys. Rev. B}} \textbf{\bibinfo{volume}{87}},
  \bibinfo{pages}{184115} (\bibinfo{year}{2013}).

\bibitem{Drautz2019-er}
\bibinfo{author}{Drautz, R.}
\newblock \bibinfo{title}{Atomic cluster expansion for accurate and
  transferable interatomic potentials}.
\newblock \emph{\bibinfo{journal}{Phys. Rev. B}} \textbf{\bibinfo{volume}{99}},
  \bibinfo{pages}{014104} (\bibinfo{year}{2019}).

\bibitem{Dusson2019-gn}
\bibinfo{author}{Dusson, G.} \emph{et~al.}
\newblock \bibinfo{title}{Atomic cluster expansion: Completeness, efficiency
  and stability}.
\newblock \emph{\bibinfo{journal}{J. Comp. Phys.}}
  \textbf{\bibinfo{volume}{454}}, \bibinfo{pages}{110946}
  (\bibinfo{year}{2022}).

\bibitem{Schutt2019}
\bibinfo{author}{Schütt, K.~T.}, \bibinfo{author}{Gastegger, M.},
  \bibinfo{author}{Tkatchenko, A.}, \bibinfo{author}{Müller, K.-R.} \&
  \bibinfo{author}{Maurer, R.~J.}
\newblock \bibinfo{title}{Unifying machine learning and quantum chemistry with
  a deep neural network for molecular wavefunctions}.
\newblock \emph{\bibinfo{journal}{Nat. Commun.}} \textbf{\bibinfo{volume}{10}},
  \bibinfo{pages}{5024} (\bibinfo{year}{2019}).

\bibitem{gasteggerDeepNeuralNetwork2020}
\bibinfo{author}{Gastegger, M.}, \bibinfo{author}{{McSloy}, A.},
  \bibinfo{author}{Luya, M.}, \bibinfo{author}{Schütt, K.~T.} \&
  \bibinfo{author}{Maurer, R.~J.}
\newblock \bibinfo{title}{A deep neural network for molecular wave functions in
  quasi-atomic minimal basis representation}.
\newblock \emph{\bibinfo{journal}{Journal of Chemical Physics}}
  \textbf{\bibinfo{volume}{153}}, \bibinfo{pages}{044123}
  (\bibinfo{year}{2020}).
\newblock \eprint{2005.06979}.

\bibitem{Hegde2017}
\bibinfo{author}{Hegde, G.} \& \bibinfo{author}{Bowen, R.~C.}
\newblock \bibinfo{title}{Machine-learned approximations to density functional
  theory hamiltonians}.
\newblock \emph{\bibinfo{journal}{Scientific Reports}}
  \textbf{\bibinfo{volume}{7}}, \bibinfo{pages}{42669} (\bibinfo{year}{2017}).

\bibitem{Bartok2010}
\bibinfo{author}{Bartók, A.~P.}, \bibinfo{author}{Payne, M.~C.},
  \bibinfo{author}{Kondor, R.} \& \bibinfo{author}{Csányi, G.}
\newblock \bibinfo{title}{Gaussian approximation potentials: The accuracy of
  quantum mechanics, without the electrons}.
\newblock \emph{\bibinfo{journal}{Phys. Rev. Lett.}}
  \textbf{\bibinfo{volume}{104}}, \bibinfo{pages}{136403}
  (\bibinfo{year}{2010}).

\bibitem{Nigam2021-eq}
\bibinfo{author}{Nigam, J.}, \bibinfo{author}{Willatt, M.~J.} \&
  \bibinfo{author}{Ceriotti, M.}
\newblock \bibinfo{title}{Equivariant representations for molecular
  hamiltonians and n-center atomic-scale properties}.
\newblock \emph{\bibinfo{journal}{J. Chem. Phys.}}
  \textbf{\bibinfo{volume}{156}}, \bibinfo{pages}{014115}
  (\bibinfo{year}{2022}).

\bibitem{UnkeNeurIPS2021}
\bibinfo{author}{Unke, O.} \emph{et~al.}
\newblock \bibinfo{title}{{SE}(3)-equivariant prediction of molecular
  wavefunctions and electronic densities}.
\newblock \emph{\bibinfo{journal}{NeurIPS}} \textbf{\bibinfo{volume}{34}},
  \bibinfo{pages}{14434--14447} (\bibinfo{year}{2021}).

\bibitem{Cances2021convergence}
\bibinfo{author}{Canc{\`e}s, E.}, \bibinfo{author}{Kemlin, G.} \&
  \bibinfo{author}{Levitt, A.}
\newblock \bibinfo{title}{Convergence analysis of direct minimization and
  self-consistent iterations}.
\newblock \emph{\bibinfo{journal}{SIAM J. Matrix Anal. Appl.}}
  \textbf{\bibinfo{volume}{42}}, \bibinfo{pages}{243--274}
  (\bibinfo{year}{2021}).

\bibitem{woodsComputingSelfconsistentField2019}
\bibinfo{author}{Woods, N.~D.}, \bibinfo{author}{Payne, M.~C.} \&
  \bibinfo{author}{Hasnip, P.~J.}
\newblock \bibinfo{title}{Computing the self-consistent field in kohn–sham
  density functional theory}.
\newblock \emph{\bibinfo{journal}{J. Phys.: Condens. Matter}}
  \textbf{\bibinfo{volume}{31}}, \bibinfo{pages}{453001}
  (\bibinfo{year}{2019}).

\bibitem{Lysogorskiy2021}
\bibinfo{author}{Lysogorskiy, Y.} \emph{et~al.}
\newblock \bibinfo{title}{Performant implementation of the atomic cluster
  expansion (pace) and application to copper and silicon}.
\newblock \emph{\bibinfo{journal}{npj Comput. Mater.}}
  \textbf{\bibinfo{volume}{7}}, \bibinfo{pages}{1--12} (\bibinfo{year}{2021}).

\bibitem{Willatt2018-ao}
\bibinfo{author}{Willatt, M.~J.}, \bibinfo{author}{Musil, F.} \&
  \bibinfo{author}{Ceriotti, M.}
\newblock \bibinfo{title}{Feature optimization for atomistic machine learning
  yields a data-driven construction of the periodic table of the elements}.
\newblock \emph{\bibinfo{journal}{Phys. Chem. Chem. Phys.}}
  \textbf{\bibinfo{volume}{20}}, \bibinfo{pages}{29661--29668}
  (\bibinfo{year}{2018}).

\bibitem{Nigam2020-re}
\bibinfo{author}{Nigam, J.}, \bibinfo{author}{Pozdnyakov, S.} \&
  \bibinfo{author}{Ceriotti, M.}
\newblock \bibinfo{title}{Recursive evaluation and iterative contraction of
  n-body equivariant features}.
\newblock \emph{\bibinfo{journal}{J. Chem. Phys.}}
  \textbf{\bibinfo{volume}{153}}, \bibinfo{pages}{121101}
  (\bibinfo{year}{2020}).

\bibitem{Drautz2020-mg}
\bibinfo{author}{Drautz, R.}
\newblock \bibinfo{title}{Atomic cluster expansion of scalar, vectorial, and
  tensorial properties including magnetism and charge transfer}.
\newblock \emph{\bibinfo{journal}{Phys. Rev. B}}
  \textbf{\bibinfo{volume}{102}}, \bibinfo{pages}{024104}
  (\bibinfo{year}{2020}).

\bibitem{Grisafi2019-in}
\bibinfo{author}{Grisafi, A.} \& \bibinfo{author}{Ceriotti, M.}
\newblock \bibinfo{title}{Incorporating long-range physics in atomic-scale
  machine learning}.
\newblock \emph{\bibinfo{journal}{J. Chem. Phys.}}
  \textbf{\bibinfo{volume}{151}}, \bibinfo{pages}{204105}
  (\bibinfo{year}{2019}).

\bibitem{Slater54}
\bibinfo{author}{Slater, J.~C.} \& \bibinfo{author}{Koster, G.~F.}
\newblock \bibinfo{title}{Simplified {LCAO} method for the periodic potential
  problem}.
\newblock \emph{\bibinfo{journal}{Phys. Rev}} \textbf{\bibinfo{volume}{94}},
  \bibinfo{pages}{1498--1524}.

\bibitem{crandall1973bifurcation}
\bibinfo{author}{Crandall, M.~G.} \& \bibinfo{author}{Rabinowitz, P.~H.}
\newblock \emph{\bibinfo{title}{Bifurcation, perturbation of simple eigenvalues
  and linearized stability}} (\bibinfo{publisher}{University of
  Wisconsin-Madison, Mathematics Research Center}, \bibinfo{year}{1973}).

\bibitem{PhysRevB.102.235130}
\bibinfo{author}{Ben~Mahmoud, C.}, \bibinfo{author}{Anelli, A.},
  \bibinfo{author}{Cs\'anyi, G.} \& \bibinfo{author}{Ceriotti, M.}
\newblock \bibinfo{title}{Learning the electronic density of states in
  condensed matter}.
\newblock \emph{\bibinfo{journal}{Phys. Rev. B}}
  \textbf{\bibinfo{volume}{102}}, \bibinfo{pages}{235130}
  (\bibinfo{year}{2020}).

\bibitem{Knoop2020}
\bibinfo{author}{Knoop, F.}, \bibinfo{author}{Purcell, T. A.~R.},
  \bibinfo{author}{Scheffler, M.} \& \bibinfo{author}{Carbogno, C.}
\newblock \bibinfo{title}{Anharmonicity measure for materials}.
\newblock \emph{\bibinfo{journal}{Phys. Rev. Materials}}
  \textbf{\bibinfo{volume}{4}}, \bibinfo{pages}{083809} (\bibinfo{year}{2020}).

\bibitem{PBE}
\bibinfo{author}{Perdew, J.~P.}, \bibinfo{author}{Burke, K.} \&
  \bibinfo{author}{Ernzerhof, M.}
\newblock \bibinfo{title}{Generalized gradient approximation made simple}.
\newblock \emph{\bibinfo{journal}{Phys. Rev. Lett.}}
  \textbf{\bibinfo{volume}{77}}, \bibinfo{pages}{3865} (\bibinfo{year}{1996}).

\bibitem{Monkhorst1976}
\bibinfo{author}{Monkhorst, H.~J.} \& \bibinfo{author}{Pack, J.~D.}
\newblock \bibinfo{title}{Special points for brillouin zone integration}.
\newblock \emph{\bibinfo{journal}{Phys. Rev. B}} \textbf{\bibinfo{volume}{13}},
  \bibinfo{pages}{5188--5192}.

\bibitem{zhou2004}
\bibinfo{author}{Zhou, X.~W.}, \bibinfo{author}{Johnson, R.~A.} \&
  \bibinfo{author}{Wadley, H. N.~G.}
\newblock \bibinfo{title}{Misfit-energy-increasing dislocations in
  vapor-deposited {CoFe}/{NiFe} multilayers}.
\newblock \emph{\bibinfo{journal}{Phys. Rev. B}} \textbf{\bibinfo{volume}{69}},
  \bibinfo{pages}{144113} (\bibinfo{year}{2004}).

\bibitem{ACE_jl}
\bibinfo{author}{Ortner, C.} \& \bibinfo{author}{{et al.}}
\newblock \bibinfo{title}{{ACE.jl}: Approximation of symmetric functions with
  polynomials and spherical harmonics} .

\end{thebibliography}
\end{document}